  \documentclass[twocolumn,aps,prb,groupedaddress]{revtex4-1}
  \usepackage{graphicx,wrapfig,amsmath,upgreek,mathtools,amsfonts,physics,multirow,color}
  \usepackage{graphicx}
  \usepackage{color}
  \usepackage{amsmath}
  \usepackage{enumitem}
  \usepackage{amssymb}
  \usepackage{hyperref}

  \newcommand{\ov}[1]{\overline{#1}}

  \newcommand{\be}{\begin{equation}}
  \newcommand{\ee}{\end{equation}}
  
  \newcommand{\bea}{\begin{eqnarray}}
  \newcommand{\eea}{\end{eqnarray}}

  \newcommand{\p}{\partial}
  
  \newcommand{\la}{\left\langle}
  \newcommand{\ra}{\right\rangle}

  \newcommand{\lp}{\left(}
  \newcommand{\rp}{\right)}
  \newcommand{\sgn}{\text{sgn}}

  \renewcommand{\Im}{{\rm \, Im\,}}
  
  \renewcommand{\vec}[1]{{\boldsymbol #1}}

\begin{document}
  \title{
  The Hierarchy of 
  Excitation Lifetimes in Two-Dimensional  
  Fermi Gases}
  \author{Patrick J Ledwith, Haoyu Guo  and Leonid Levitov} 
  \affiliation{Massachusetts Institute of Technology, Cambridge, Massachusetts 02139, USA}

  \begin{abstract}
  Momentum-conserving quasiparticle collisions in two-dimensional Fermi gases give rise to a large family of exceptionally long-lived excitation modes. The lifetimes of these modes exceed by a factor $(T_F/T)^2\gg 1$ the conventional Landau Fermi-liquid lifetimes $\tau\sim T_F/T^2$.  The long-lived modes have a distinct angular structure, taking the form of $\cos m\theta$ and $\sin m\theta$ with odd $m$ values for a circular Fermi surface, with relaxation rate dependence on $m$ of the form $m^4\log m$, valid at not-too-large $m$. In contrast, the even-$m$ harmonics feature conventional lifetimes with a weak $m$ dependence. The long-time dynamics, governed by the long-lived modes, takes the form of angular (super)diffusion over the Fermi surface. Altogether, this leads to unusual long-time memory effects, defining an intriguing transport regime that lies between the conventional ballistic and hydrodynamic regimes.
  \end{abstract}
  \maketitle
\tableofcontents

  \section{The long- and short-lived modes: angular structure 
  and dynamics}

Describing degenerate two-dimensional (2D) electrons in terms of Fermi surface geometry that varies in space and time is a powerful approach that links the ideas of one-dimensional bosonization with Landau quasiparticles and collective modes\cite{Haldane1994,Houghton1993,CastroNeto1994}. This approach interprets the low-energy excitations of the Fermi liquid as fluctuations of the shape of the Fermi
surface, treating these fluctuations as bosonic fields. Recently, these ideas were successfully applied to the problem of the Fractional Quantum Hall effect\cite{Golkar2016,Nguyen2018} and to Fermi liquids with spin-orbit coupling\cite{Kumar2017}. Here we show that similar ideas help to gain new insight into the long-standing problem of quasiparticle lifetimes and angular relaxation in 2D Fermi liquids.

Fermi-liquid theory 
describes elementary excitations in degenerate Fermi gases as free-fermion quasiparticles with finite lifetimes governed by two-body collisions\cite{SmithJensen,LifshitzPitaevskii,Reif}. In three-dimensional (3D) systems, the low-temperature collision rate $\gamma\sim T^2/T_F$
  sets the timescale
  \be\label{eq:tau*}
  \tau_*=\gamma^{-1}
  \ee
  that separates two fundamentally different transport regimes:  ballistic at short times $t<\tau_*$, and hydrodynamic at longer times $t>\tau_*$. Hydrodynamic transport is governed by the modes associated with the quantities conserved due to microscopic conservation laws (energy, momentum and particle number), whereas the memory about all nonconserved quantities is quickly erased at times $t\gtrsim\tau_*$.

This picture, well established theoretically and thoroughly tested experimentally in 3D Fermi liquids, \cite{Abrikosov_Khalatnikov,baym_pethick} 
must undergo a substantial revision in two dimensions. The new behavior arises due to the interplay between kinematics of elastic collisions and fermion exclusion, which render the head-on  collisions the dominant mechanism of angular relaxation\cite{laikhtman_headon,gurzhi_headon,molenkamp_headon}. A new family of emergent conserved quantities, resulting from such dynamics,
  gives rise to a new hierarchy of time scales. This hierarchy defines a new ``tomographic" regime that lies in between the conventional ballistic and hydrodynamic regimes. 
Dynamics in the tomographic regime 
feature strong directional memory and slow angular relaxation, which lead to scale-dependent viscosity and peculiar nonlocal 
effects at times $t> \tau_*$\cite{ledwith2017,ledwith2017b}.

\begin{figure}[t]
\includegraphics[width=0.99\columnwidth]{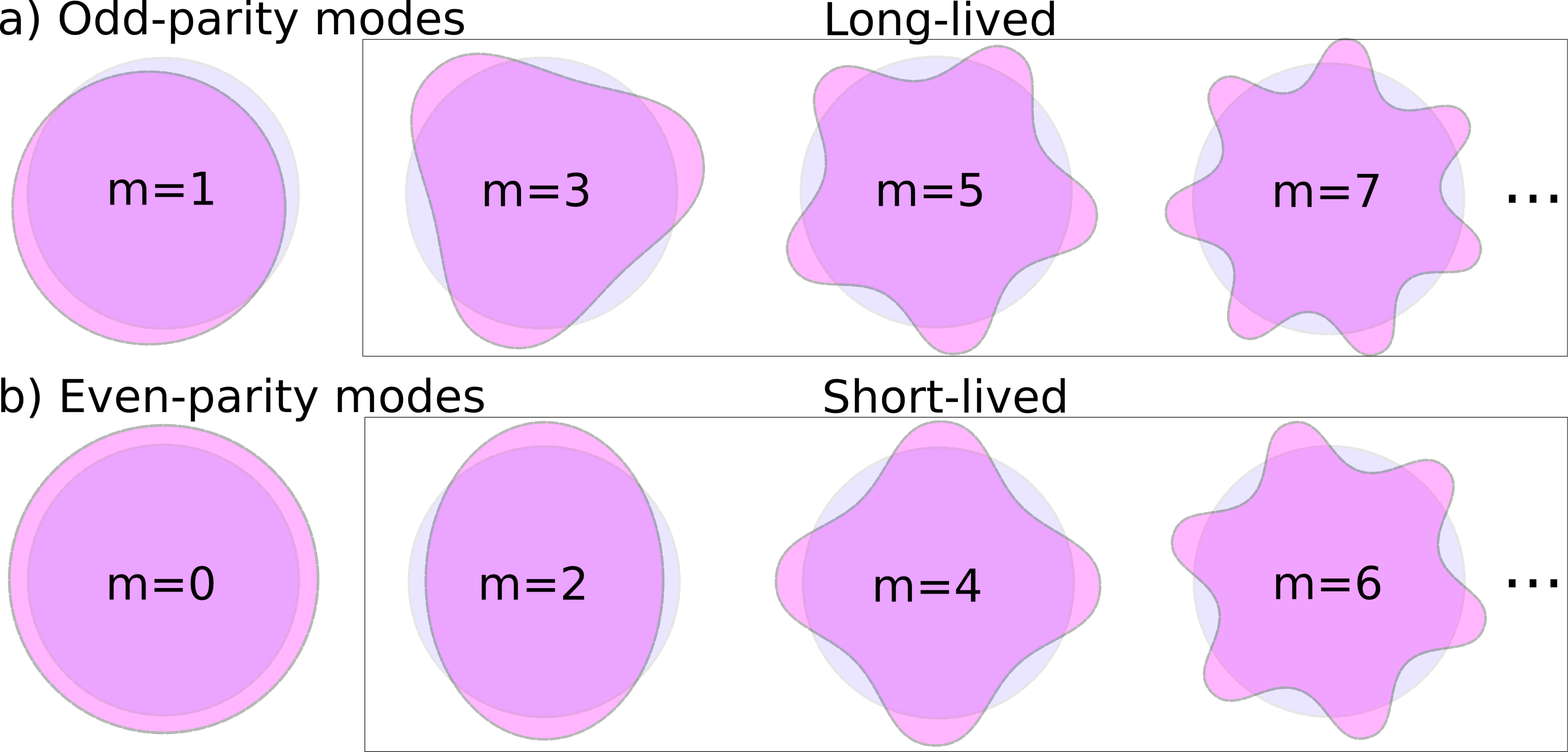}
\caption{The angular dependence of different perturbations of the Fermi surface. The odd-$m$ modes have odd parity under inversion symmetry $\vec k\to -\vec k$ and are long-lived. The even-$m$ modes have even parity and are short-lived. The modes with $m=0$ and $1$ do not relax, since they represent the zero-modes of the collision operator originating from particle number and momentum conservation, respectively.
}
 \label{fig1}
\vspace{-5mm}
\end{figure}

The new hierarchy and the resulting anomalous kinetics can be captured most naturally by representing
excitations as perturbations of the Fermi surface shape with different angular structure. This representation is particularly useful when different harmonics have widely varying decay rates, as is the case here. Here we consider a circular Fermi surface, describing perturbations by cylindrical harmonics $\cos m\theta$ and $\sin m\theta$ with integer $m$ values, as illustrated in Fig.\ref{fig1}. As we will see, the even-$m$ harmonics retain the `normal' decay rates $\gamma_{m\,{\rm even}}\sim T^2/T_F$ ($T\ll T_F$), whereas the odd-$m$ harmonics exhibit exceptionally small decay rates 
\be\label{eq:odd/even_ratio}
\gamma_{m\,{\rm odd}}\sim (T/T_F)^2\gamma_{m\,{\rm even}}
.
\ee
The large difference between the even-$m$ and odd-$m$ rates gives rise to a multiscale dynamics, in which some degrees of freedom undergo fast equilibration, whereas other degrees of freedom remain excited and dynamically active for a long time after being activated.

Harmonics $e^{im\theta}$ form a complete set of functions that can be used to analyze time evolution of perturbations with any angular structure. For example, a 
quasiparticle with momentum $\vec k=k_F(\cos\theta_0,\sin\theta_0)$ is represented by a bump on the Fermi surface located at $\theta=\theta_0$. Approximating the bump by a delta function $\alpha \delta(\theta-\theta_0)$, we can write the time evolution as
\be
\label{eq:time_evolution}
\delta f(\theta,t)=\sum_{m=-\infty}^\infty \alpha e^{im(\theta-\theta_0)-\gamma_m t}
.
\ee
The resulting relaxation dynamics is of a multiscale character, since the even-$m$ harmonics decay at the conventional $\gamma_m\sim T^2/T_F$ rates, whereas the odd-$m$ harmonics with not-too-high $m$ 
decay considerably more slowly. This strong $m$ dependence is in contrast with the behavior in 3D Fermi liquids, where the analysis of lifetimes in the angular harmonics basis\cite{brooker1972} 
gives relaxation at the conventional rates for all low-order harmonics.

Physically, the reason for abnormally slow decay of the odd-$m$ harmonics lies in that these harmonics relax through {\it many repeated collisions}, taking place at times $t\gg \tau_*$. In contrast,  the even-$m$ harmonics relax at the one-collision timescale, $t\sim\tau_*$. Different timescales arise despite that all quasiparticle  collisions occur at a rate $\gamma\sim T^2/T_F$.
Indeed, we will see that kinematic constraints and fermion exclusion, acting together, restrict possible scattering processes to
near head-on collisions; the near head-on nature of collisions makes them much more effective in relaxing even harmonics than odd harmonics.  Even harmonics therefore relax at the one-collision timescale, whereas odd harmonics require many collisions to relax.

To clarify the hierarchy of relaxation timescales, we analyze the
scaling $\gamma_m$ vs. $m$, which turns out to be quite interesting.
For low-lying excitations, we find
\begin{align}\nonumber
&\gamma_{m\,{\rm even}}\sim \frac{T^2}{T_F} \log m
,\quad
\gamma_{m\,{\rm odd}}\sim \frac{T^4}{T_F^3} m^4\log m
,\quad
\\ \label{eq:gamma_even_odd}
& 1<m\lesssim m_{\rm max}=\sqrt{T_F/T},
\end{align}
valid up to numerical prefactors that depend on the two-body interaction strength. The odd harmonics are considerably more long-lived than the even ones, so long as $m<m_{\rm max}$. At $m$ exceeding $m_{\rm max}$ the even/odd asymmetry disappears and the standard Landau scaling $\gamma_m\sim T^2/T_F$ is recovered.
\begin{figure}[tb]
  \centering
  \includegraphics[width = 0.5\textwidth]{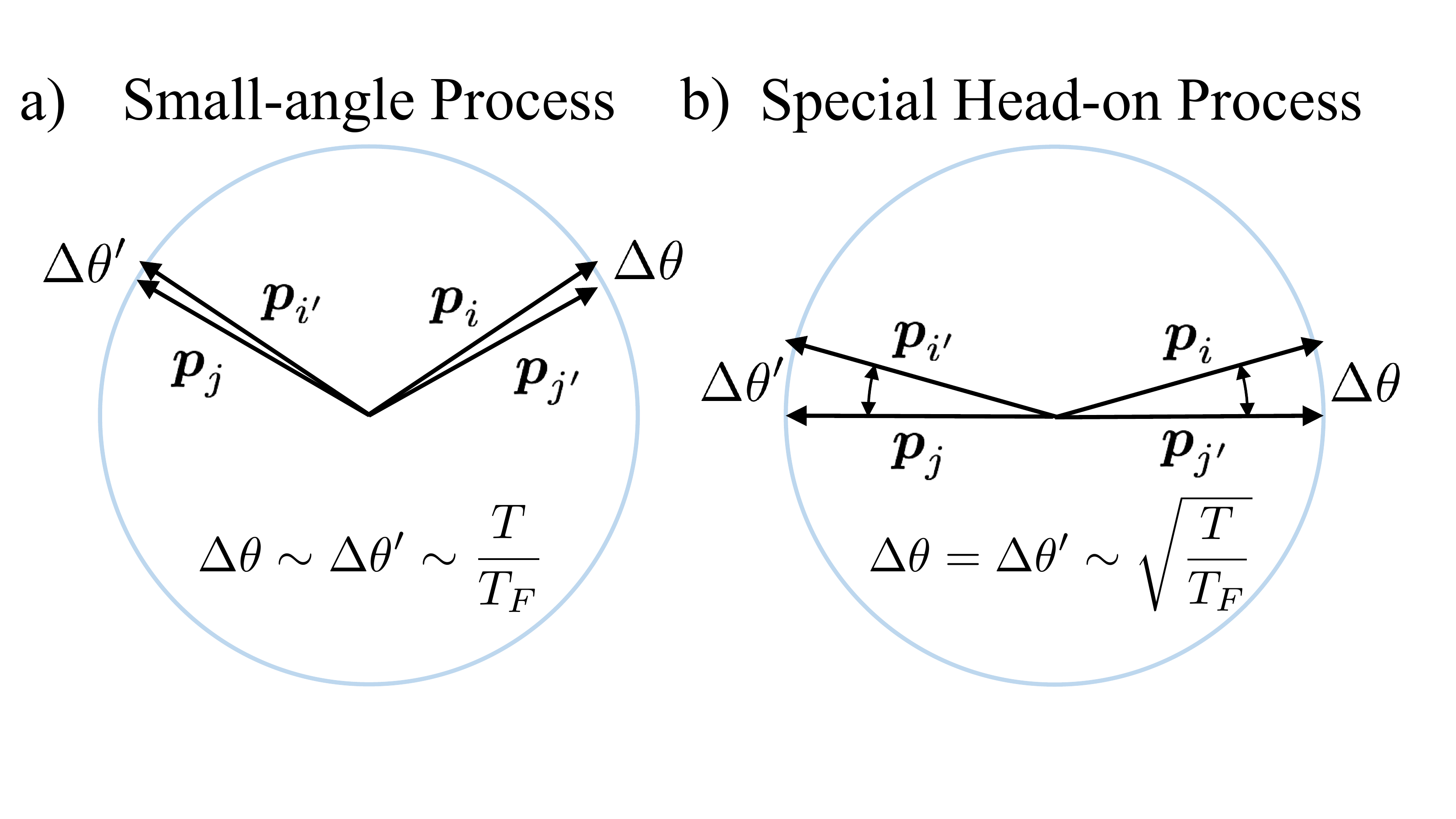}
\caption{
  Schematics of two collision types: a small-angle process (a) and a ``special head-on'' process (b). The small-angle process leads to conventional diffusion on the Fermi surface, but its contribution is subleading to the special head-on 
process which leads to ``superdiffusion''.
Particle momenta $\vec p_i, \vec p_j$ are ingoing and $\vec p_{i'}, \vec p_{j'}$ are outgoing. In the a) process, the collision is nearly forward (with exchange) and particles both hop by small, uncorrelated, angles $\Delta \theta \sim \Delta \theta' \sim \frac{T}{T_F}$.
In the b) process the particles are additionally nearly head-on and nearly-collinear,  leading to enhanced scattering angles that are identical at leading order in $T/T_F\ll 1$: 
$\Delta\theta = \Delta \theta' \sim (T/T_F)^{1/2}$. We call the b) process a special head-on process: it dominates the odd-m rates and leads to the $\gamma_m \sim m^4$ scaling.}
\label{fig:specialprocess}
\end{figure}

The $m^4$ scaling of the odd-$m$ rates arises due to a combination of several effects.  The collisions that relax odd harmonics are small-angle processes and near-head-on processes; 
examples are shown in Fig.\ref{fig:specialprocess} a and b. The process of particular importance for us is a 
special type of near head-on process 
shown in Fig.\ref{fig:specialprocess}b  that is also small angle. We will see that both small-angle and near-head-on processes can be regarded as small angular steps for an odd-parity distribution.
Normally, such dynamics would be described by angular diffusion---Brownian random walk on the Fermi surface---which would result in the rates $\gamma_m=Dm^2$, where $D$ is the angular diffusion coefficient. This is indeed the case for typical small angle collisions such as the one in Fig.\ref{fig:specialprocess}, as well as typical near head-on collisions.

  However, it turns out that in our problem angular relaxation is dominated by 
collisions with nontrivial two-particle correlations of angular displacements that do not 
result in simple angular diffusion. 
These are the special head-on processes in which all four momenta (the two ingoing and the two outgoing) are near-collinear and opposite to each other as depicted in Fig. \ref{fig:specialprocess}b. These processes lead to enhanced angular step sizes as compared to other small-angle or near-head-on processes  ($\delta\theta\sim (T/T_F)^{1/2}$ vs. $\delta\theta\sim T/T_F$), which makes them dominate angular relaxation. At the same time, as we will see, the angular steps of the two colliding particles are equal to each other at leading order in $T/T_F$. The enhanced angular stepsize and enhanced correlations, acting together, generate the unusual $m^4$ scaling. This behavior will be discussed in more detail in Sec.\ref{sec:discuss_superdiffusion}.

The long-time dynamics at $t>\tau_*$ can be viewed as an angular diffusion process through which the excitation gradually spreads over the entire Fermi surface. The form of the diffusion operator is mandated by the $m$ dependence of the rates in Eq.\eqref{eq:gamma_even_odd}, giving (with logarithmic accuracy)
\be\label{eq:superdiffusion}
(\p_t+D\p_\theta^p)\delta f(\theta,t)=0
,\qquad
p=4
,
\ee
where the distribution obeys the odd-parity condition $\delta f(\pi+\theta,t)=-\delta f(\theta,t)$. The dynamics is distinct from a normal diffusive Brownian walk in angles associated with the rates $\gamma_m=Dm^2$, with $D$ the angular diffusion coefficient.
The anomalous angular diffusion, described by Eq.\eqref{eq:superdiffusion}, originates from multiple repeated head-on collisions, arising due to enhanced angular stepsize and correlations of angular steps in
the special head-on collisions 
as discussed above. 
We will refer to this behavior as {\it superdiffusion,} described 
by a square of the Laplacian $\p_\theta^4$, Eq.\eqref{eq:superdiffusion}. 

As a side remark, ``superdiffusion'' is often used in the literature as a name for anomalous diffusion described by
$(\p_t-D\nabla^p)n(x,t)=0$ with the exponent $p<2$, whereas the case $p>2$ goes under the name ``subdiffusion.'' Our choice for the name ``superdiffusion'', while not entirely conventional, is meant to reflect the ``anomalously fast'' relaxation rates as compared to the ``normal'' angular diffusion rates  $\gamma_m=Dm^2$.

It is interesting to mention that angular superdiffusion of the form reminiscent of our Eq.\eqref{eq:superdiffusion} has appeared, a long time ago, in an entirely different context. In 1970's,  Gurevich and Laikhtman analyzed energy and momentum
transport in fluids, which at low enough temperatures 
is dominated by near-collinear scattering between acoustic phonons\cite{gurevich1975,gurevich1976,gurevich1979}.
Such processes lead to fast thermalization for each given direction,
establishing, on a relatively short time scale, an angle-dependent temperature. The latter evolves on a longer time scale through angular superdiffusion with $p=4$, described by an equation similar to Eq.\eqref{eq:superdiffusion} on a 2D sphere in 3D momentum space.

Perhaps the most direct manifestation of angular diffusion can be seen through spreading of a collimated beam of particles injected into the system. Eq.\eqref{eq:superdiffusion}, combined with the tomographic 
space-time evolution\cite{ledwith2017b}, predicts gradual  decollimation of the injected beam, spreading as
\be
\delta \theta\sim t^{1/p},\quad
\delta \theta\sim x^{2/p}
\ee
where $t$ is time and $x$ is the distance measured along the beam line, related through the scaling $x\sim t^{1/2}$ originating from tomographic dynamics\cite{ledwith2017b}. This and other related effects will be discussed in more detail in Sec.\ref{sec:Conclusions}.

A  quick note on notations before we 
proceed to the technical discussion. Throughout the paper, unless specified otherwise, we adopt 
units $k_B=\hbar=1$, restoring correct dimensions in the final results. In these units, we have $\varepsilon_F=T_F$ for Fermi energy and $p_F=k_F$ for Fermi momentum. The notation $T_F$ will be used most of the time; $k_F$ and $p_F$ will be used 
interchangeably. In discussing two-body scattering we will sometimes refer, for brevity, to different particle momentum states as ``particles''.

The outline of the paper is as follows. In Sec.\ref{sec:kinetic_eqn}, we will introduce the
Boltzmann kinetic equation formalism for fermion collisions.
In Sec.\ref{sec:The_Eigenvalue_problem}, we will represent the problem of finding
the decay rates $\gamma_m$ as a linear eigenvalue problem of the linearized collision operator.
We will construct a Hilbert space of
excitations, and show that the collision operator is Hermitian with a suitable inner product. In Sec.\ref{sec:kinematic_constraints}, we introduce a parameterization of the configuration space that accounts for the kinematic constraints in a way convenient for subsequent analysis.
In Secs.\ref{sec:The_Eigenvalue_problem},\ref{sec:kinematic_constraints}, we will compute the low-lying eigenvalues to zeroth order in $T/T_F$, 
recovering the conventional Fermi-liquid result for even-$m$ harmonics, 
but finding a vanishing scattering rate for odd-$m$ harmonics. In Sec.\ref{sec:next_steps}
we discuss the strategy for developing perturbation theory in $T/T_F$ in order to compute the odd-$m$ relaxation rates.
In Sec.\ref{sec:perturbation_theory}, which is central for this work, we will set up a Rayleigh-Schrodinger-like perturbation theory formalism, using 
$T/T_F\ll 1$ as a small parameter. In Sec.\ref{sec:matrix_element_evaluation}
we will evaluate the matrix elements that appear in the perturbation analysis, for simplicity ignoring collisions with small momentum transfer $\vec q \approx 0$ and half of the terms in the integrand. 
We will see that the matrix elements 
exhibit log divergences at
$q\sim 2k_F$
that are naturally cut off by a finite energy transfer $\omega \sim T$.
This analysis predicts scaling of relaxation rates with $T$ and $m$, as given in Eq.\eqref{eq:gamma_even_odd}, up to numerical factors that are established in subsequent sections. In Sec.\ref{sec:discuss_superdiffusion} we pause to discuss the physical picture, in particular the correlations between angular displacements  of scattering particles
that underpin the $m^4$ and $T^4$ scaling, as well as implications of the latter for angular relaxation (superdiffusive behavior). Next, in Secs.\ref{sec:duality},\ref{sec:full_calculation} we patch the analysis of Sec.\ref{sec:matrix_element_evaluation} to account for the contributions of
forward scattering $q
\approx 0$
and the other half of the integrand. In Sec.\ref{sec:duality}, we will invoke a geometric duality and reflection symmetry to
show that the final result does not change except for a factor of two.
In Sec.\ref{sec:full_calculation}, we redo the calculation with the collisions with $\vec q \approx 0$ and the entire integrand included from the start. In Sec.\ref{sec:Conclusions}, we will summarize the results and discuss 
possible experimental implications.

\section{The kinetic equation approach. Why collision operator?}
\label{sec:kinetic_eqn}

In order to examine the long-lived states, we will develop an approach based on 
the kinetic equation
\be\label{eq:kinetic_eqn}
(\partial_t + \vec v\cdot \vec \nabla)f(\vec p) = I[f(\vec p)],
\ee
where the collision operator $I[f(\vec p)]$ describes two-body collisions of quasiparticles
with energies near the Fermi level. We will focus, exclusively, on perturbations about the low-temperature state, $0<T\ll T_F$. The quantity of primary interest for us will be the collision operator $I[f(\vec p)]$ linearized in weak perturbations from the equilibrium state. The eigenmodes and eigenvalues of this operator 
describe different excitations and their decay rates, respectively.

We note in passing 
that higher-body collisions give rise to a smaller collision rate.  For example,  
the standard phase-space counting argument shows that three-body collisions have a base rate of $T^4/T_F^3$ (arising from five energy integrals subject to one constraint) which is smaller than the 
odd-parity and even-parity rates, Eq.\eqref{eq:gamma_even_odd}, found from the two-body collision processes. We also note that the kinetic equation for quasiparticles in a Fermi liquid includes the Landau mean-field interaction term that modifies the $\vec v\cdot \vec \nabla$ term in Eq.\eqref{eq:kinetic_eqn}.

For two-body collisions, $I[f(\vec p)]$ is expressed as a difference of rates  of 
the ``gain" and ``loss" processes that populate and depopulate a state with momentum $\vec p_i$,
\begin{equation} \label{inandout}
   I[f(\vec p_i)] = \int \frac{d^2 p_j d^2 p_{i'} d^2 p_{j'}}{(2\pi)^6} (W_{i'j' \to ij} - W_{ij \to i'j'}),
\end{equation}
where $j$, $i'$ and $j'$ label the other particle states involved in the collision.  The transition rates are given by Fermi's golden rule as
\begin{align}\nonumber
&  W_{ij \to i'j'} = \frac{2\pi}{\hbar}|V|^2 f_i f_j (1-f_{i'})(1-f_{j'})
\\
& \times \delta \left(\sideset{ }{'}
\sum_\alpha \varepsilon_\alpha \right)  (2\pi)^2 \delta^{(2)} \left( \sideset{ }{'}\sum_\alpha \vec p_{\alpha} \right),
  \label{goldenrate}
\end{align}
where $V$ is a shorthand notation for the interaction matrix element which will be defined and discussed below.
The primed summations in Eq.\eqref{goldenrate} denote a difference between ingoing and outgoing quantities,
\begin{equation}
  \sideset{ }{'}\sum_\alpha A_\alpha = A_i + A_j - A_{i'} - A_{j'},
  \label{primesum}
\end{equation}
so that the delta functions implement energy and momentum conservation.

In subsequent sections, we present a detailed analysis of the quantity $I[f(\vec p)]$ linearized in the deviations from equilibrium, and use it to describe different types of excitations and their decay rates. However, there are several aspects of the collision operator approach that must be discussed first.

One has to do with the properties of the interaction matrix element $V$ in Eq.\eqref{goldenrate}.
For spinless particles, the matrix element $V$ is given by the (in general, screened) two-body interaction
\be
U(\vec r-\vec r')=\int \frac{d^2p}{(2\pi)^2} U_{\vec p} e^{i\vec p(\vec r-\vec r')}
\ee
antisymmetrized under fermion exchange: 
\be 
V=\la \vec p_i,\vec p_j|U|\vec p_{i'},\vec p_{j'}\ra=U_{\vec p_i-\vec p_{i'}}-U_{\vec p_i-\vec p_{j'}}
.
\ee
For spin-$1/2$ particles and spin-independent interaction $U_{\vec p}$, we have
\be\label{eq:V=U-U}
|V|^2=|U_{\vec p_i-\vec p_{i'}}|^2+|U_{\vec p_i-\vec p_{j'}}|^2 +|U_{\vec p_i-\vec p_{i'}}-U_{\vec p_i-\vec p_{j'}}|^2
\ee
where the first two terms describe scattering of two particles with opposite spins $\sigma_i=\sigma_{i'}\ne \sigma_j=\sigma_{j'}$ and $\sigma_i=\sigma_{j'}\ne \sigma_j=\sigma_{i'}$, whereas the last term describes scattering of particles with equal spins.  We assume spin-unpolarized distributions, described by probability $1/2$ for each spin component.

The details of the dependence of $V$ on particle momenta, summarized here for  completeness, will not matter in our analysis.  Instead, there is one specific value of $\abs{V}^2$ that will appear, corresponding to special head-on processes in which momenta are near-collinear and opposite to each other, 
such as the one depicted in Fig.\ref{fig:specialprocess}b. This gives
\be\label{specialV_spinless}
 \abs{V_*}^2=|U_0 - U_{2k_F}|^2
 \ee
 for spinless particles, and
 \be\label{specialV_spinful}
  \abs{V_*}^2=\abs{U_0}^2 + \abs{U_{2k_F}}^2 +\abs{U_0 - U_{2k_F}}^2
  \ee
  for spin-$1/2$ particles.

Another question of interest has to do with the choice of theoretical framework to analyze excitation lifetimes. Indeed, at this point, the educated reader might be wondering about
the relation between the present approach and the conventional analysis of excitation lifetimes in Fermi liquids based on the Green's function selfenergy  calculations\cite{coleman2015,galitskii1958,morel1962}.
The latter approach, as is well known, predicts decay rates  scaling with temperature as $\gamma\sim T^2/T_F$ in both 3D and 2D Fermi liquids. Furthermore, in 2D systems 
the rates exhibit additional enhancement by a log factor $\log \frac{T_F}{T}$.
\cite{chaplik1971,hodges1971,bloom1975,giuliani1982,zheng1996,menashe1996,chubukov2003}
The selfenergy approach is therefore conspicuously unaware of the existence of the long-lived odd-parity excitations.

The resolution of this conundrum lies in the peculiar multiscale character of relaxation dynamics in our system. Indeed, it is usually taken for granted that there is a single timescale that characterizes decay for all low-energy excitations. However, this is very much untrue for 
2D, since the odd-parity modes have lifetimes that are 
considerably longer than those of the even-parity modes. The conventional selfenergy approach is not well suited for such a situation, since
selfenergy is the quantity which is most sensitive to the fastest decay pathways.

As discussed above, the new behavior arises because the predominantly head-on collisions give rise to slow angular relaxation. The corresponding characteristic times are those of many repeated collisions rather than one-collision. Because the selfenergy is dominated by the fast-decaying modes, it does not capture the contribution of slow-decaying modes, which remain `hidden' in the selfenergy calculation.

\section{The eigenvalue problem for the linearized collision operator}
\label{sec:The_Eigenvalue_problem}

Our first step will be to linearize the collision operator $I[f(\vec p)]$ in a small deviation from the equilibrium distribution,
$f(\vec p) = f^{(0)}(\varepsilon_{\vec p}) + \delta f(\vec p)$. We will use the standard ansatz
\begin{align}\nonumber
&\delta f(\vec p) = F (\varepsilon)\eta(\vec p),
\quad
\\
&F(\varepsilon) = -\pdv{f^{(0)}}{\varepsilon} = \beta f^{(0)}(1-f^{(0)})
,
  \label{linearization}
\end{align}
where $f^{(0)}(\varepsilon) = (e^{\beta \varepsilon} + 1)^{-1}$ is the Fermi distribution function at temperature $T = \beta^{-1}$ with energy $\varepsilon$ measured relative to the Fermi energy $\varepsilon_F$.
The quantity $\eta(\vec{p})$ 
can be viewed as a small momentum-dependent perturbation to chemical potential. Linearizing the gain and loss terms in Eq.\eqref{inandout} in $\delta f(\vec p)$ parameterized through $\eta(\vec{p})$, and simplifying the result, brings the collision operator to a compact form
\begin{widetext}
\begin{equation}\label{eq:I[f]compact}
  \begin{aligned}
   I[\eta (\vec p_i)]  = & -\beta \int \frac{d^2 p_j d^2 p_{i'} d^2 p_{j'}}{(2\pi)^6} \frac{2\pi}{\hbar} |V|^2 f^{(0)}_if^{(0)}_j
 (1-f^{(0)}_{i'})(1-f^{(0)}_{j'})
\\
& \times
    \delta \left(\sideset{ }{'}\sum_\alpha \varepsilon_\alpha \right)  (2\pi)^2 \delta^{(2)} \left( \sideset{ }{'}\sum_\alpha \vec{p}_{\alpha} \right)\sideset{ }{'}\sum_\alpha \eta_\alpha 
.
  \end{aligned}
\end{equation}
\end{widetext}
Here the primed sums denote the difference of the ingoing and outgoing quantities as in Eq.\eqref{primesum}. For instance,
\be\label{eq:sum'eta}
\sideset{ }{'}\sum_\alpha \eta_\alpha=\eta_i+\eta_j-\eta_{i'}-\eta_{j'}
.
\ee
From now on we will drop the superscript $(0)$ on equilibrium Fermi functions and will adopt a shorthand notation $\eta_\alpha=\eta(\vec p_\alpha)$. To simplify notation, we will be using a single integral symbol, as in Eq.\eqref{eq:I[f]compact}, to denote multiple integration.

We wish to examine eigenmodes $\delta f(\vec p)$ such that
\begin{equation}
I[\eta(\vec p)] = -\gamma \delta f(\vec p)=-\gamma F(\epsilon) \eta(\vec p).
  \label{fevp}
\end{equation}
In what follows it will be convenient to rescale the collision operator and define a new  operator
\begin{equation}
   L[\eta(\vec p)] = F^{-1} I[\eta(\vec p)]
   ,
  \label{Ldef}
\end{equation}
which transforms the eigenvalue problem to the form
\begin{equation}
  L[\eta(\vec p)] = -\gamma_m  \eta(\vec p)
  \label{phievp}
\end{equation}
 that we will analyze below. By rotational invariance, we can look for solutions of the form $\delta f(\vec{p}) = g(p) e^{i m \theta_p}$ for integer $m$, as depicted in Fig.\ref{fig1}, and label the eigenvalues $-\gamma_m$.

We first note that there is a small set of eigenmodes with zero eigenvalues, two for $m=0$ and two more for $m=1$:
\be
\eta(\vec p)=1
,\quad
\eta(\vec p)=\varepsilon
,\quad
\eta(\vec p)=p_x
,\quad
\eta(\vec p)=p_y
.
\ee
These are nothing but the zero modes of the collision operator
originating from conservation of the particle number, energy and momentum in two-body collisions.

\begin{figure}[t]
\includegraphics[width=0.99\columnwidth]{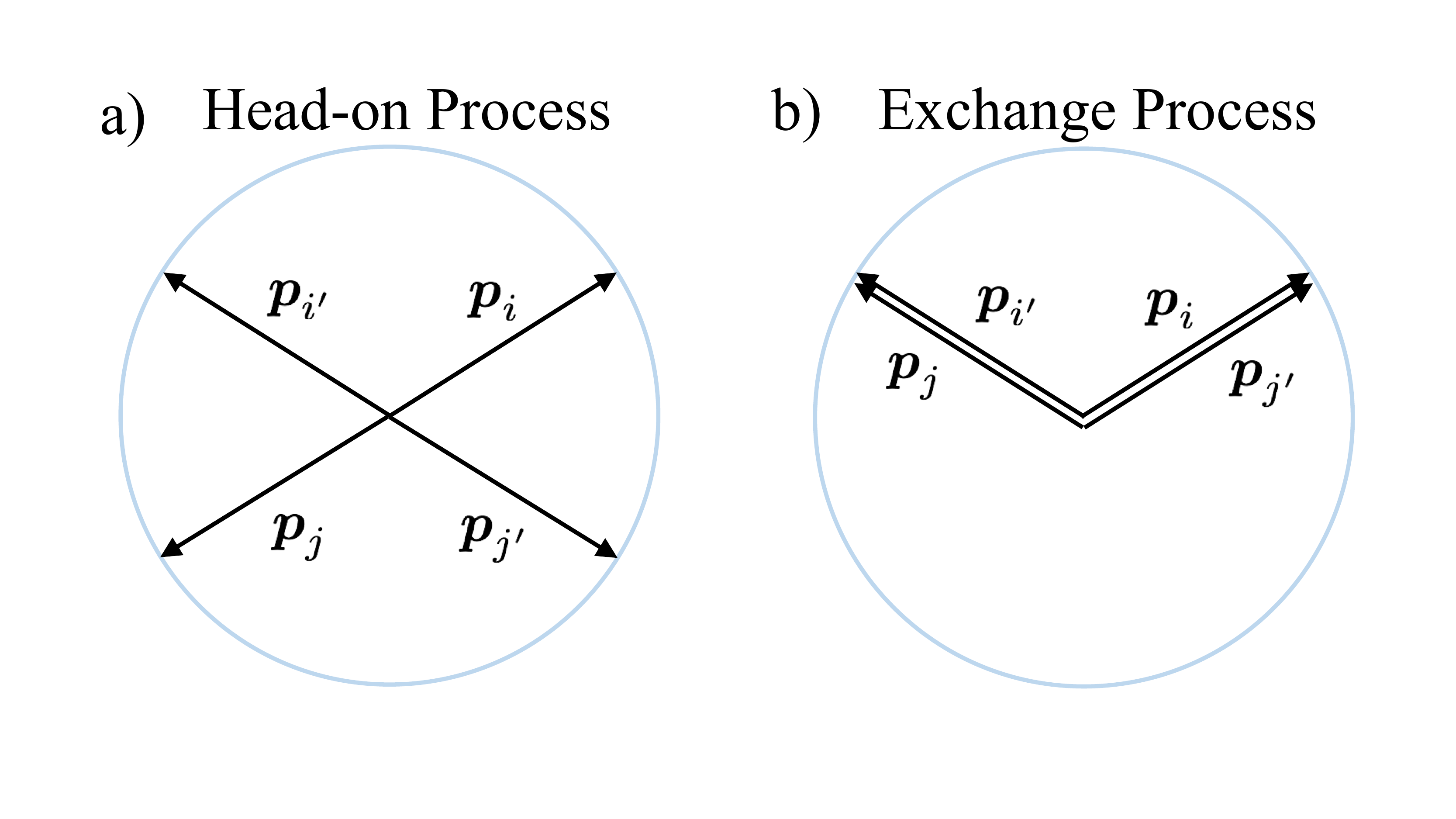}
\caption{Schematic of the possible two-body collisions of quasiparticles at a sharp Fermi surface.  Shown are two possible scattering processes, arising due to the kinematic and fermion exclusion restrictions: a) The incident particle momenta $\vec p_i$ and $\vec p_j$ are head-on, the outgoing momenta $\vec p_{i'}$ and $\vec p_{j'}$ are head-out. In this case the scattering angles are unconstrained. b) The incident particle momenta are at a generic angle, and the scattering is forward (up to possible  exchange). In this case the outgoing  particle momenta are constrained to be at the same angles as the incident momenta. 
}
 \label{fig2}
\end{figure}

To motivate the analysis of other eigenmodes 
on which 
we embark below,
it is instructive to consider a leading-order behavior at low temperatures $T\ll T_F$. At such temperatures, the quantities $F$ in Eq.\eqref{linearization} and $ f^{(0)}_if^{(0)}_j(1-f^{(0)}_{i'})(1-f^{(0)}_{j'})$ in Eq.\eqref{eq:I[f]compact} behave as delta functions centered at the Fermi level, pinning all four energies $\varepsilon_\alpha$, $\alpha=i,j,i',j'$ to $\varepsilon=\varepsilon_F$. Combining these restrictions with the kinematic constraints due to momentum conservation, $\vec p_i+\vec p_j=\vec p_{i'}+\vec p_{j'}$, we find that the only allowed scattering processes leading to angular relaxation over the Fermi surface are the head-on collisions for which the momenta $\vec p_\alpha$ satisfy\cite{ledwith2017}
\be\label{eq:headon}
\vec p_i=-\vec p_j
,\quad
\vec p_{i'}=-\vec p_{j'}
,
\ee
as pictured in Fig.\ref{fig2} (a).
In this case the odd-$m$ angular harmonics obey
\be
e^{im\theta_i} +e^{im\theta_j}=0,\qquad e^{im\theta_{i'}} +e^{i m\theta_{j'}}=0
.
\ee
These relations ensure that the quantity in Eq.\eqref{eq:sum'eta} vanishes, giving zero eigenvalues at leading order in $T\ll T_F$ for all the modes with odd $m$. At the same time, as discussed in more detail below, the even-$m$ modes have nonzero eigenvalues of the ``normal" scale $\gamma\sim T^2/T_F$.
This conclusion is unaffected by the presence of two other solutions of the kinematic constraints, $\vec p_i=\vec p_{i'}$, $\vec p_j=\vec p_{j'}$ and $\vec p_i=\vec p_{j'}$, $\vec p_j=\vec p_{i'}$. These solutions describe forward particle scattering with possible exchange, as illustrated in Fig.\ref{fig2} (b), 
a process that does not contribute to angular relaxation.

We will see that, while the odd-$m$ eigenvalues do vanish at leading order in small $T/T_F\ll 1$, they are nonzero at a higher order.
To determine 
these eigenvalues, we therefore need to go beyond the conventional Sommerfeld approximation that treats the thermally broadened Fermi surface as a delta-function energy shell.
Below we bring the expression for collision integral to the form that will facilitate this analysis, and then proceed to develop a systematic perturbation theory in the $T/ T_F$ parameter.

\section{Resolving kinematic constraints}
\label{sec:kinematic_constraints}

We start with writing the integrals over energies and momenta in a way that makes the temperature dependence  in $L[\eta(\vec p)]$ more apparent.  
We split the energy and momentum delta functions by introducing integrals over the energy and momentum transferred between colliding particles $i$ and $j$: 
\begin{widetext}
\begin{align}
&  \delta\lp\sideset{ }{'}\sum_\alpha \varepsilon_\alpha\rp =  \int d\omega \, \delta(\varepsilon_i - \varepsilon_{i'} - \omega) \delta(\varepsilon_j - \varepsilon_{j'} + \omega).
  \label{energysplit}
  \\
 & \delta^{(2)}\lp\sideset{ }{'}\sum_\alpha \vec p_\alpha\rp = \int d^2q \, \delta^{(2)}(\vec{p}_i-\vec{p}_{i'}-\vec{q})
 \delta^{(2)}(\vec{p}_j-\vec{p}_{j'}+\vec{q}),
  \label{momentumsplit}
\end{align}
\end{widetext}
and integrate over the outgoing momenta $\vec p_{i'}$, $\vec p_{j'}$ so that we are just left with an integral over the momentum transfer $\vec{q}$.

Throughout the paper we will use a parabolic band dispersion
\be
\varepsilon = \frac{p^2}{2m_*} - \varepsilon_F
,
\ee
where, following our convention, the energy is measured from $\varepsilon_F$.
The parabolic model will be convenient because it simplifies algebra without affecting the general applicability of our conclusions, so long as temperature is small compared to Fermi energy, $T\ll \varepsilon_F$. Indeed, for any band dispersion with cylindrical symmetry, the only relevant parameter that controls the behavior of states
 sufficiently close to the Fermi level 
 is the effective mass $m_* = p_F/v_F$.

We 
therefore decompose
\begin{align}\nonumber
\int d^2 p_j &= m_* \int_{-\infty}^\infty d \varepsilon_j \oint d\theta_j
,\quad
\\
\int d^2 q &= \int_0^\infty qdq \oint d\theta_q
,
  \label{energyanglesplit}
\end{align}
noting that because the equilibrium Fermi functions are exponentially decaying away from the Fermi level, they restrict energies to $\abs{\varepsilon_\alpha} \sim T \ll \varepsilon_F$. The integral over energy difference from the Fermi level can therefore be continued to $-\infty$.
Integrating the energy delta functions over the angles $\theta_q$ and $\theta_j$ gives:
\begin{widetext}
\begin{equation}
  \begin{aligned}
    \int d\theta_q \,\delta(\varepsilon_i-\varepsilon_{i'} - \omega)  &= \int d\theta_q \, \delta\left(v_iq\cos (\theta_i-\theta_q)
 - q^2/2m_* - \omega \right) = \sum_{s_1 = \pm1} \frac{1}{v_i q \abs{\sin (\theta_i-\theta_q)
}}, \\
    \int d\theta_j \,\delta(\varepsilon_j-\varepsilon_{j'} + \omega) &= \int d\theta_j \,\delta\left(-v_jq\cos (\theta_j-\theta_q)
- q^2/2m_* + \omega \right) = \sum_{s_2 = \pm1} \frac{1}{v_j q \abs{\sin  (\theta_j-\theta_q)
}}.
\end{aligned}
  \label{energydeltas}
\end{equation}
\end{widetext}
The asymmetry between these expressions, with $d\theta_q$ appearing in place of $d\theta_i$, is due to the fact that we are not integrating over $\theta_i$.

\begin{figure}[t]
\includegraphics[width=0.99\columnwidth]{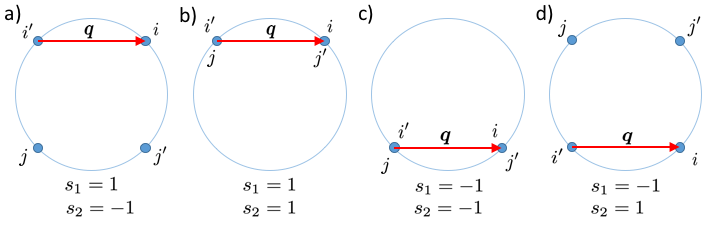}
\caption{Schematics of possible collision processes allowed by kinematic constraints, with momentum transfer $\vec q$, which is chosen to be horizontal. 
The four cases shown correspond to different values of $s_{1,2}=\pm 1$.
The sign $s_1$ determines whether particles $i$ and $i'$ have momenta above $(+)$ or below $(-)$ the $q$ axis.  The sign $s_2$ controls, in a similar manner, the momenta
of particles $j$ and $j'$.  If $s_1 = -s_2$ particles scatter through a head-on process, relaxing the even-$m$ harmonics;
if $s_1 = s_2$ the collision is of an exchange type and the distribution remains unchanged.
}
 \label{fig3}
\end{figure}

In what follows it will be convenient to measure all the angles relative to the $\vec q$ direction, making use of the rotational invariance of the problem. We will use $\theta_\alpha-\theta_q$ as new angular variables and,
unless stated otherwise, will use $\theta_\alpha$ as a shorthand notation for $\theta_\alpha-\theta_q$. E.g. in Eq.\eqref{energydeltas}, we will have $\frac1{
\sin\theta_i}$ and $\frac1{
\sin\theta_j}$ instead of $\frac1{\sin  (\theta_i-\theta_q)}$ and $\frac1{\sin  (\theta_j-\theta_q)}$.

The sign factors $s_{1,2}=\pm$ appearing in Eq.\eqref{energydeltas}, which are defined by
\begin{equation}
  \begin{aligned}
    s_1 & = \sgn(\sin \theta_i) = \sgn(\sin \theta_{i'}), \\
    s_2 & = \sgn(\sin \theta_j) = \sgn(\sin \theta_{j'} ),
  \end{aligned}
  \label{s1s2}
\end{equation}
label the roots of the arguments of the delta functions.
The angles $\theta_\alpha$, found by resolving the delta function constraints in Eq.\eqref{energydeltas}, are given by the closed-form expressions: 
\begin{equation}
  \begin{aligned}
    \cos \theta_i & = +\frac{q}{2p_i} + \frac{\omega}{v_iq}, \quad
    \cos \theta_j  = -\frac{q}{2p_j} + \frac{\omega}{v_jq}, \\
    \cos \theta_{i'} & = -\frac{q}{2p_{i'}} + \frac{\omega}{v_{i'}q}, \quad
  \cos \theta_{j'}  = +\frac{q}{2p_{j'}} + \frac{\omega}{v_{j'}q}.
  \end{aligned}
  \label{angles}
\end{equation}
Possible collision processes, described by different combinations of $s_1$ and $s_2$, are shown in Fig.\ref{fig3}. For illustration, all the states are taken on the $T=0$ Fermi surface such that $|\vec p_\alpha|=p_F$ and $\omega=0$. In contrast, the angles given in Eq.\eqref{angles} are exact.

The relations in Eq.\eqref{energydeltas} can now be used to simplify   the collision operator, Eq.\ref{eq:I[f]compact}.
Using the quantity $L[\eta(\vec p)] = F^{-1} I[\eta(\vec p)]$ introduced above, and rescaling energies as
\be  \label{eq:energies_u_w}
u_\alpha = \varepsilon_\alpha/T, \quad
w = \omega/T
,
\ee
we obtain
\begin{widetext}
\begin{align}
  L[\eta(\vec{p}_i)] = \frac{-2 m_*T^2}{\hbar(2\pi)^3f_i(1-f_i)}\sum_{s_1, s_2} \int du_j d w f_if_j
 (1-f_{i'})(1-f_{j'}) \int \frac{dq}{qv_iv_j\abs{\sin \theta_i \sin \theta_j 
  }}
 |V|^2 \sideset{ }{'}\sum_\alpha \eta(\vec{p}_\alpha)
  ,
  \label{qomega}
\end{align}
\end{widetext}
an expression that exhibits the ``natural scale'' 
$T^2/T_F$ of $L$.  The rescaled energies $u_j$ and $w$ are integrated from $-\infty$ to $\infty$ as in \eqref{energyanglesplit}.  The bounds on the $q$ integration in Eq.\eqref{qomega} are fixed by the requirement that the angles in \eqref{angles} satisfy $\abs{\cos \theta_\alpha} \leq 1$.
The bounds are therefore set by the $q$ values where $\abs{\sin \theta_i \sin \theta_j} = 0$. To zeroth order in $T/T_F$ these are $q=0$ and $q=2k_F$.
These bounds will be analyzed more explicitly in later sections.

The form of Eq.\eqref{qomega} is convenient for the purpose of our analysis, since energy integration is separated from the $q$ integration. The latter, 
in our choice of variables, serves as proxy for angular integration. 
However, since the angles $\theta_\alpha$ do depend on the energies $\varepsilon_\alpha$ (through $\omega$ and $v_\alpha$), the collision operator exhibits a nontrivial interplay between the angular and energy dynamics. Accounting for this interplay is key for understanding the kinetics due to head-on collisions and, eventually, obtaining a correct estimate for the odd-$m$ rates. 
This will be the main subject of our interest below.

We now estimate the rates $\gamma_m$ to lowest order $(T^2)$. This will provide 
a simple application of $\eqref{qomega}$ 
and will help to clarify the unique role of the head-on processes.
At low temperature, $T\ll T_F$, the expression is dominated by processes where all four 
energies are 
on the Fermi level.  In this limit, neglecting in Eqs.\eqref{angles} the energy transfer $\omega\sim T$ compared to $q^2/2m_*$, we find that the angles obey
\be
\cos\theta_i=-\cos\theta_j=-\cos\theta_{i'}=\cos\theta_{j'}.
\ee
This condition means that the collisions are head-on  ($s_i = -s_2$, Fig. \ref{fig3}a,d) or forward with possible exchange ($s_1 = s_2$, Fig. \ref{fig3}b,c).  The latter possibility leads to $\eta_i + \eta_j = \eta_{i'} + \eta_{j'}$ implying $L[\eta(\vec p_i)] = 0$.  We therefore only need to consider the head-on collisions
  $\vec p_i = -\vec p_j$, $\vec p_{i'} = -\vec p_{j'}$.
For $\eta(\vec p) = e^{im\theta_p}$ we find
\begin{equation}
  \sum_{s_1,s_2} \sideset{ }{'}\sum_\alpha \eta(\vec p_\alpha) = e^{im\theta_i}\begin{cases}
    4(1-\cos2m\theta) & \text{$m$ even}, \\
    0 & \text{$m$ odd},
  \end{cases}
  \label{even_m_etas}
\end{equation}
where we defined the angle $\theta$ which equals $\pi$ minus the scattering angle:
\be
\cos \theta = x = \frac{q}{2k_F}.
\label{xandtheta}
\ee
The dimensionless quantities $\theta$ and $x$ will be convenient for our analysis of the odd-m rates as well.  For even $m$, we can write the integration over $q$ as
\begin{equation}
  \begin{aligned}
    & \int dq \frac{4(1-\cos2m\theta)}{qv_iv_j\abs{\sin\theta_i \sin\theta_j}} \abs{V}^2, \\
  & \approx \frac{4}{v^2} \int_0^{\pi/2} \frac{d\theta}{\cos\theta\sin\theta}\abs{V}^2 (1-\cos2m\theta) .
  \end{aligned}
  \label{even}
\end{equation}
The apparent divergence in the denominator at $\theta=0,\pi/2$ is cut off logarithmically at $\theta \sim 1/m$ and $ \pi/2-\theta \sim 1/m$ by the $(1-\cos 2m\theta)$ factor.  We can use this to argue for
\begin{equation}
  \gamma_{m \,\rm{ even}} \sim \frac{T^2}{T_F}\abs{V_*}^2 \log m,
  \label{gamma_meven}
\end{equation}
as while $\eta(\vec p) = e^{im\theta_p}$ is not an eigenvector of $L$, the angular portion of it must be, and this is what leads to the $\log m$ enhancement.  The matrix element $\abs{V_*}^2$ arises because for $\theta \to 0$ we have $(\vec p_i - \vec p_{i'}) \to 2k_F$ and $(\vec p_i - \vec p_{j'}) \to 0$ and vice-versa for $\theta \to \pi/2$.  Thus both limits reproduce the expressions for $V_*$ in Eqs.\eqref{specialV_spinless},\eqref{specialV_spinful} and correspond to a special head-on process such as the one in Fig.\ref{fig:specialprocess}b.  While here the special head-on processes dominate due to log-enhancement, for the odd-harmonics they will also be favored due to an enhancement $m^2 \to m^4$.

If $m$ is larger than $T_F/T$, the integrand is cut off by energy transfer instead and we obtain \begin{equation}
  \gamma_{m > T_F/T} \sim \frac{T^2}{T_F}\log \frac{T_F}{T}
,
  \label{large_m}
\end{equation}
a result familiar in 2D Fermi liquids\cite{chaplik1971,hodges1971,bloom1975,giuliani1982,zheng1996,menashe1996,chubukov2003}.


The above estimate 
is 
good for the eigenmodes 
with an even-$m$ angular dependence.
However, if the perturbation $\eta(\vec p)$ is an odd-$m$ harmonic that is slowly varying in momentum magnitude, e.g. $\eta(\vec{p}) = e^{im\theta_p}$ for odd $m$, then the contributions of head-on collisions vanish.  In this case, the only collisions that can lead to a nonzero value of $\gamma_m$
are collisions with momenta
slightly off the Fermi level, which are suppressed at low temperature by some power of $T/T_F$.

We therefore conclude that the spectrum of the collision integral has multiple timescales. One is the conventional timescale due to $T^2/T_F$ collision rate for even harmonics. The other, longer, timescale is due to
the odd-harmonics relaxation rates
that we expect to scale with a higher power of temperature. We will find a scaling  $T^4/T_F^3$ with a prefactor that behaves as $m^4\ln m$ at not-too-high $m$ values.  The rest of this paper will focus on determining the odd-$m$ relaxation rates, and thus from now on $m$ will always represent some odd integer.

\section{Strategy for odd-$m$ rates}
\label{sec:next_steps}

Here we pause for a moment to reflect upon the results so far and to discuss subsequent steps.
We start with identifying the hurdles that are encountered in developing perturbation theory, and then discuss how those are resolved.
\begin{itemize}
\item
One unusual aspect of the problem at hand is the complex structure of the configuration space, parameterized by momenta of the three particle states $j$, $i'$ and $j'$, which are subject to the kinematic constraints due to energy and momentum conservation. Six momentum components and three delta functions translate into a three-dimensional integration in the collision integral, Eq.\eqref{eq:I[f]compact}.
\item Besides being three-dimensional, the configuration space for two-body scattering has a fairly complicated structure:  For each of the four participating particle states $i$, $j$, $i'$, $j'$, all the action is happening in a thin shell centered on the 2D Fermi sphere broadened by $\delta p=T/v_F\ll k_F$, whereas the inner states are blocked by fermion exclusion. The kinematic constraints due to momentum conservation are encoded through the angles defined in Eq.\eqref{angles}.
\item Energy and momentum transferred between the particles in the collisions result in energy steps and angular steps that are coupled in a nontrivial way. Indeed, because of the $\omega/q$ dependence in Eq.\eqref{angles},  small values $\omega\sim T$ may not always translate into small values for the angular steps. As a result, our analysis, in which we treat $\omega$ as a small perturbation, will take a very different route away from $q=0$ and near $q=0$.
\item Last but not least, we encounter unexpected cancellations in perturbation theory not just at leading order but also beyond leading order. To understand the general structure of perturbation theory, and to handle these cancellations, we introduce a Hilbert space that describes various perturbations $\eta(\vec p)$ in a unified way. We develop perturbation theory using the linear operator framework and the quantum-mechanical Dirac notation, which is a not a common approach in statistical mechanics problems but is indispensable in this case due to the complex nature of the problem.
\end{itemize}
To resolve these issues we proceed as follows. We use the parameterization of the configuration space through the angles and nondimensionalized energies,  defined in Eq.\eqref{angles} and Eq.\eqref{eq:energies_u_w}. In the next section we define the Hilbert space of perturbations $\eta(\vec p)$ and use it to set up perturbation theory separately for each harmonic order $m$ value. Then we perform perturbation analysis away from the region $q\approx 0$, discuss cancellations and determine the leading-order dependence of the eigenvalues. Then we show that the behavior near  $q\approx 0$ is related to that away from $q\approx 0$ by a suitably defined duality transformation. We use the duality argument to refine the analysis and to show that, up to a combinatorial factor, the results found away from $q\approx 0$ remain unaltered and have a completely general validity.

\section{
Eigenvalue perturbation theory at low temperatures}
\label{sec:perturbation_theory}

We need to go beyond lowest order in temperature in order to compute the odd-$m$
relaxation rates.  To do this, we
develop eigenvalue perturbation theory with the small parameter
\begin{equation}
  \delta_T = T/T_F\ll 1.
  \label{taudefn}
\end{equation}
We will start  with a general discussion of how this expansion works in practice, postponing the details to the next section.
The odd-$m$ rate must come from processes
slightly off the Fermi level so that the combination $\eta_i + \eta_j - \eta_{i'} - \eta_{j'}$ does not vanish.  We can expand $\eta_\alpha$'s around zero deviation from the Fermi level, and obtain a power series in $\varepsilon_\alpha/\varepsilon_F = \delta_T u_\alpha$, for $\alpha = i,j,i',j'$.

The power series in $\delta_Tu_\alpha$ is integrated against the Fermi functions.  The integration produces no further temperature dependence other than that of $\delta_T$, as all quantities are appropriately nondimensionalized. We therefore obtain $I[\eta]$ as a power series in $\delta_T$.  We translate this expansion to an expansion of $\gamma_m$ using eigenvalue perturbation theory, and compute the corrections to the lowest order zero mode $\eta(\vec p) = e^{im\theta_{\vec p}}$.

To make the arguments in this section 
more transparent we introduce a 
compact notation for the angular and radial parts of the measure
\begin{widetext}
  \begin{equation}
    \begin{aligned}
     d\nu_{iji'j'}  &= \frac{m_*}{(2\pi)^3\hbar}\frac{dq}{qv_iv_j\abs{\sin \theta_i  \sin \theta_j }} \abs{V}^2, \\
   d\mu_{ji'j'}  &= \frac{1}{f_i(1-f_i)} du_j dw f_if_j(1-f_{i'})(1-f_{j'}) \\
      & = \frac{1}{f_i(1-f_i)} du_j du_{i'}du_{j'}\delta(u_i+u_j-u_{i'}-u_{j'})  f_if_j(1-f_{i'})(1-f_{j'}).
    \end{aligned}
\label{nudefn}
\end{equation}
\end{widetext}
The angular measure $d\nu_{iji'j'}$ is (manifestly) symmetric under exchanging $i$ and $j$; it is also symmetric under ``reversing the time arrow" by
swapping
ingoing and outgoing states due to momentum conservation in the direction perpendicular to $\vec q$.  We can write $L$ as
\begin{equation}
  L[\eta(\vec{p}_i)] = -T^2\int d\mu_{ji'j'} \sum_{s_1, s_2} \int d\nu_{iji'j'} \sideset{ }{'}\sum_\alpha \eta(\vec{p}_\alpha)
  .
  \label{eq:schematicL}
\end{equation}
Below we consider the space of perturbations $\eta(\vec p)$, taken separately for each harmonic order $m$. It is natural to endow this space of functions with a Hilbert space structure, by defining the inner product as
\begin{equation}
  \braket{\eta'}{\eta} = \frac{1}{2\pi m_*} \int d^2 p \overline{\eta'(\vec p)}F(p) \eta(\vec p).
  \label{innerproduct}
\end{equation}

Importantly, $L$ is Hermitian with respect to this inner product. To show this, we consider the matrix element $\bra{\eta'}L\ket{\eta}$.  The energy dependence in the factor of $F$ in the inner product is 
canceled out by the $1/f_i(1-f_i)$ in $d\mu_{ji'j'}$, leaving a residual factor of $\beta$.  We write $\int d^2 p_i = 2\pi m_*T \int du_i$, where the integral over $\theta_i$ only gives a factor of $2\pi$ due to rotation symmetry.  This factor cancels the remaining part of the inner product normalization.  We then obtain
  \begin{equation}
    \bra {\eta'} L \ket \eta = -T^2 \int d\mu_{iji'j'} \sum_{s_1, s_2} \int d\nu_{iji'j'} \ov{\eta'}_i \sideset{ }{'}\sum_\alpha \eta_\alpha,
\label{eq:L_eta_eta'}
\end{equation}
  where the energy integration measure is now given by
  \begin{equation}
    \begin{aligned}
     & d\mu _{iji'j'}  = du_i du_j dw f_if_j(1-f_{i'})(1-f_{j'}),\\
      & = du_i du_j du_{i'}du_{j'} \delta(u_i+u_j-u_{i'}-u_{j'})
\\
& \times f_if_j(1-f_{i'})(1-f_{j'}).
  \end{aligned}
    \label{energymeasure}
  \end{equation}
  Crucially, this measure is invariant under exchanging $i$ and $j$ as well as 
under
swapping ingoing and outgoing states because the equilibrium Fermi functions satisfy $1-f(-u) = f(u)$.  Since this is true for $d\nu_{iji'j'}$ as well, we can symmetrize $\ov{\eta'}_i$ with respect to exchanging the ingoing states $i$ and $j$ and antisymmetrize it with respect to swapping
ingoing and outgoing states. This symmetry property holds because the quantity 
$\sum'_\alpha \eta(\vec p_\alpha)$ is even and odd under these exchanges and swaps. 
After symmetrization, the matrix element in Eq.\eqref{eq:L_eta_eta'} is brought to a manifestly symmetric form
  \begin{align}\nonumber
 & \bra {\eta'} L \ket \eta = -\frac{1}{4}T^2 \int d\mu_{iji'j'} \sum_{s_1, s_2} \int d\nu_{iji'j'}
\\
&
\times \sideset{ }{'}\sum_\alpha \ov{\eta'}_\alpha \sideset{ }{'}\sum_\alpha \eta_\alpha = \ov{\bra \eta L \ket {\eta'}},
    \label{schematicmelt}
  \end{align}
and so $L$ is Hermitian.
Additionally, if we plug in $\eta' = \eta$, we obtain a non-positive expression and so $L$ is negative semidefinite, as expected on general grounds given that we want $\gamma_m$ to be real and non-negative.

Before we proceed further, we comment on how the expansion in $\delta_T$ is 
formally accomplished. In Eqs.(23) we replace $\omega/v_i q$ with $\delta_T w p_i/2 q$, write momenta and velocities as
\be\label{eq:p_i,v_i_expansion}
p_i=p_F+\varepsilon_i/v_F+...
,\quad
v_i=v_F+\varepsilon_i/p_F
,
\ee
and rescale $\varepsilon_i$ as in Eq.\eqref{eq:energies_u_w} above. The part of $L$ that needs to be expanded, as usual in perturbation theory, involves matrix elements with fixed $\ket {\eta}$ and  $\ket {\eta'}$. The part of $L$ which involves the integration measure
\be
\int d\mu_{iji'j'} ... =\int du_i du_jdw f_i f_j(1-f_{i'})(1-f_{j'}) ... ,
\ee
only includes properly rescaled quantities, and thus does not generate any factors of $\delta_T$. Expansion in $L[\eta(\vec p)]$ mainly comes from perturbing $\vec p$ values at which $\eta_\alpha$ ($\alpha=i,j,i',j'$) is evaluated.  Technically, the Jacobian part $v_iv_j |\sin\theta_i\sin\theta_j|$ also needs to be expanded, however, we will see that this expansion will generate terms subleading in $\delta_T$. This is so because the combination $\sum'_\alpha\eta_\alpha$ vanishes at zeroth order, and therefore, unless it is expanded to higher order, the resulting expression will vanish as well.


We write the eigenvector and the generalized eigenvalue
as power-law series expansion in our small parameter $\delta_T$, Eq.\eqref{taudefn},
\be
\begin{aligned}
\ket{\eta}& = \ket{\eta^{(0)}} + \ket{\eta^{(1)}} + \ket{\eta^{(2)}} + \cdots
\\
\gamma_m &= \gamma_m^{(0)} + \gamma_m^{(1)} + \gamma_m^{(2)} + \cdots
,
\end{aligned}
\ee
 with temperature dependence $\ket{\eta^{(n)}}\propto \delta_T^n$
and $\gamma^{(n)}\propto \delta_T^n T^2/T_F$.
Similarly, we expand the collision operator
\be\label{eq:L_expansion}
L = L^{(0)} + L^{(1)} +  L^{(2)} + \cdots.
\ee
This is done by accounting for energy-dependent changes in momenta, velocities and angles in Eq.\eqref{eq:p_i,v_i_expansion} and Eq.\eqref{angles}, as well as for the changes in $\eta_\alpha=\eta(\vec p_\alpha)$ due to $\vec p_\alpha$ dependence on $\omega$ and $p_\alpha$ in Eq.\eqref{angles} (as we will see, the latter contributions will be the most important in our analysis). Similar to $\gamma^{(n)}$, the quantities $L^{(n)}$ are order $n+2$ in $T/T_F$ because of the base rate of $T^2/T_F$.


The lowest order eigenvector is \begin{equation}
  \ket{\eta^{(0)}} = \ket{1} := 1e^{im\theta}
,
  \label{1state}
\end{equation}
and it is a zero mode to lowest order, $\gamma_m^{(0)} = 0$.
We also define a vector
\begin{equation}
  \ket{u} := \frac{u}{2} e^{im\theta}
,
  \label{varepsilonstate}
\end{equation}
where $u/2 = u(\vec{p})/2 = \varepsilon(\vec{p})/2T$ represents momentum magnitude variation near Fermi level. The vector $\ket{1}$ is normalized, $\bra{1}\ket{1} = 1$. The vector $\ket{u}$, which is not normalized, includes a prefactor $1/2$ introduced to avoid numerical factors 2 and 4 in various expressions below.
Here the notation ``$:=$'' is used, in analogy with Dirac quantum mechanics, to identify the ``quantum states" and the corresponding ``wavefunctions". 

The quantity $\ket{u}$ represents a small odd-$m$ harmonic temperature fluctuation and has a central role in our analysis.  In particular, we will show that $\ket{\eta^{(1)}} \propto \ket{u}$.
The importance of this mode reflects the complications of odd parity angular relaxation in that it is no longer possible to disentangle angular and radial relaxation.  In particular, momentum conservation forces every angular step to be paired with a radial step as long as the collisions are not perfectly head-on.  Since we will have to tackle collisions that are not head-on in order to allow odd parity modes to relax, we also need to worry about coupling to radial modes.

Properly accounting for the interplay between radial and angular displacements is important also because, as we will find below, ignoring this coupling leads to the $m= \pm 1$ modes not being conserved, while reinstating it repairs momentum conservation.  Furthermore, moving beyond lowest order allows for violations of the approximate particle-hole symmetry at the Fermi level, $u \to -u$, and as a result the state $\ket{u}$ will appear in the series in the powers of $\delta_T$, describing perturbation correction to $\ket{1}$,
despite the two states having different parity under $u \to -u$.

We first discuss the structure of $L^{(1)}\ket{1}$.  As discussed above, expanding the combination $\eta_i + \eta_j - \eta_{i'} - \eta_{j'}$ to linear order in $\delta_T$ gives a linear combination of $\delta_T u_\alpha$ for $\alpha = i,j,i',j'$.  These factors pass through the $q$ integral and 
are integrated over energies as 
\[
\int d\mu_{ji'j'} u_\alpha
.
\]
Importantly, there is a simple relation between
these quantities taken for different $\alpha=i,j,i',j'$. Namely,
all $u_{\alpha}$ factors generate an identical dependence on $u_i$ up to an overall prefactor. This can be verified by using the permutation symmetry of the second expression for $d\mu_{ji'j'}$.
Specifically, a direct evaluation of the integral
(for details, see \hyperref[appendix:integrals]{Appendix}) 
shows that
\begin{align}\nonumber
&\int d\mu_{ji'j'} u_{i'}   = \int d\mu_{ji'j'} u_{j'} = -\int d\mu_{ji'j'} u_j
\\
& = \frac{1}{3}u_i\int d\mu_{ji'j'}
=\frac{1}{6}u_i\lp \pi^2+u_i^2\rp.
\label{L1|1>}
\end{align}
Hence, $L^{(1)}\ket{1}$ is proportional to $u_i \int d\mu_{ji'j'}$ with a prefactor that depends on temperature as $\delta_T$ times the base rate $T^2/T_F$.

Crucially, same situation occurs if we compute $L^{(0)}\ket{u}$, wherein $L^{(0)}$ is the part of $L$ zero order in $\delta_T$, i.e. taken without accounting for the radial and angular displacements in $\vec p_\alpha$ proportional to $\delta_T$.
The factors of $u_\alpha$ arise in this case just from $\eta_\alpha = u_\alpha e^{im\theta}$ rather than a linear expansion within $L$.
We therefore have
\begin{equation}
 L^{(1)}\ket{1}=\lambda
L^{(0)}\ket{u} 
  \label{firstevec}
\end{equation}
with a numerical factor $\lambda$ that depends on temperature as $\delta_T$. The precise value of $\lambda$ will not matter for our discussion.  This relationship will simplify the perturbation theory we develop in the rest of the section such that in the end we will only need to compute matrix elements of $L$ involving $\ket{1}$ and $\ket{u}$.
We also note
that the expressions $L^{(1)} \ket{1}$ and $L^{0}\ket{u}$ actually have matching combinations of $ u_\alpha$ even prior to integration,
however this property is not required to deduce \eqref{firstevec}.

We note parenthetically that the above $u_i$ dependence $u_i(\pi^2 + u_i^2)$ is not precisely correct due to log-divergences in the integration over $q$ for $q \to 0$ and $q \to 2k_F$.  These divergences are cut off in an energy dependent way leading to terms like $\log(\varepsilon_F/\omega)$.  These additional contributions to the integrand in \eqref{L1|1>} generate additional $u_i$ dependence that may differ between $L^{(1)}\ket{1}$ and $L^{(0)}\ket{u}$. The final result holds, however, if we note that the relevant energy differences are of order $T$. Indeed, writing $\log(\varepsilon_F/\omega) = \log(\varepsilon_F/T) + \log(T/\omega) \approx \log(\varepsilon_F/T)$ shows that the actual energy dependence of the logarithm does not matter very much as long as $T\ll\varepsilon_F$. We will therefore ignore these contributions.

Having done this groundwork we are well equipped to discuss the perturbation calculation of the odd-$m$ eigenvalues.
Naively, at the lowest nonvanishing order in $\delta_T$ the answer for the eigenvalue $\gamma_m$ is given by the diagonal matrix element $\bra{1} L^{(1)}\ket{1}$. However, since $L^{(1)}\ket{1}$ is odd under $u \to -u$ (see Eq.\eqref{L1|1>}),
we have $\gamma_m^{(1)} = \bra{1} L^{(1)} \ket{1} = 0$ in addition to $\gamma_m^{(0)}=0$.  A second-order calculation is then necessary, at which point one must consider the influence of first-order eigenvector corrections in addition to the diagonal contribution $\bra{1} L^{(2)} \ket{1}$.

This second-order calculation, in general quite tedious, can be simplified considerably by taking into account that the lowest eigenvalue is much smaller than all other eigenvalues. The latter is true because 
the unperturbed lowest eigenvalue is zero, whereas other eigenvalues are on the order of the base rate $T^2/T_F$. The relative smallness of the lowest eigenvalue can be exploited to estimate it at lowest nonvanishing order in $\delta_T$ as
\begin{equation}
  -\gamma_m^{(2)} = \bra{1}L^{(2)}\ket{1} - \bra{1}L^{(1)} \frac1{ \tilde L^{(0)}
} L^{(1)} \ket{1}
,
  \label{generalpert}
\end{equation}
where the tilde over $L^{(0)}$ indicates that the operator $\tilde L^{(0)}$ is restricted to the subspace of vectors orthogonal to $\ket{1}$. We note in passing that, since we just showed that $L^{(1)}\ket{1}$ is orthogonal to $\ket{1}$, the expression above can be simplified by dropping the tilde.
The interplay between the two terms in Eq.\eqref{generalpert}, which are of the same order in powers of temperature but have opposite signs, is going to be important in our discussion below.

In terms of the Rayleigh-Schroedinger perturbation theory the expression in Eq.\eqref{generalpert} represents a sum of the diagonal and off-diagonal contributions 
arising at second-order perturbation theory. The second contribution can be written as $-\bra{\eta^{(1)}}L^{(0)}\ket{\eta^{(1)}}$, where $\ket{\eta^{(1)}}=(L^{(0)})^{-1}L^{(1)}\ket{1}$ is a correction to eigenvector $\ket{1}$ first-order in $\delta_T$. Comparing to 
Eq.\eqref{firstevec} above, we see that the vector $\ket{\eta^{(1)}}$ is nothing but $\ket{u}$:
\be\label{eq:eigenvector_correction_eta1}
\ket{\eta^{(1)}}=-\lambda\ket{u}
,\quad
\lambda\sim\delta_T
.
\ee
In the Rayleigh-Schroedinger perturbation theory, the lowest eigenvalue shift 
due to eigenvector change is of a negative sign, which can be interpreted as the effect of level repulsion in a quantum system. The negative sign of this contribution will be important in our discussion below, as it will cancel (partially or fully) the positive contribution due to the first, diagonal term. This cancellation will help to maintain zero values for the $m=\pm1$ eigenvalues as required by momentum conservation.


In order to derive these results, we write $L$ in a block form by decomposing the Hilbert space as $V=\ket{1}\oplus \tilde V$, where $\tilde V$ is the subspace orthogonal to $\ket{1}$:
\be
L=\lp \begin{array}{cc}
z_0 & \mu^\dagger\\ \mu & \tilde L
\end{array}
\rp
\ee
where $\tilde L$ is $L$ restricted to the subspace $\tilde V$,  $z_0=\bra{1}L\ket{1}$, the vector $\mu$ is given by $L\ket{1}$ and
$\dagger$ indicates Hermitian transpose (where complex conjugation is needed only for $e^{im\theta}$ factors).
We consider the resolvent $R(z)=1/(z-L)$, computing it with the help of the standard recipe for inverting $(n+1)\times(n+1)$ block matrices of the form\cite{block_inversion}
\be
M=\lp \begin{array}{cc}
c & \vec b^\dagger\\ \vec b & A
\end{array}
\rp
,
\ee
where $A$ is an $n\times n$ matrix and $\vec b$ is an $n$-component vector. The inverse $M^{-1}$ equals
\be
M^{-1}=\lp \begin{array}{cc}
k^{-1}& -k^{-1}\vec b^\dagger A^{-1}\\ -k^{-1}A^{-1}\vec b & A^{-1}+k^{-1}A^{-1} \vec b\vec b^\dagger A^{-1}
\end{array}
\rp
\ee
where 
$k=c-\vec b^\dagger A^{-1}\vec b
$.
This result can be used to write the resolvent of $L$ in a closed form. We will be interested, in particular, in the matrix element $R_{11}(z)=\bra{1}\frac1{z-L}\ket{1}$ 
which is given by
\be
R_{11}(z)=\frac1{k}=\frac1{z-z_0-\mu'\frac1{z-\tilde L}\mu}
.
\ee
The poles of the resolvent coincide with the eigenvalue spectrum. The equation for the poles, after plugging in $z_0=\bra{1}L\ket{1}$ and $\mu=L\ket{1}$, becomes
\be
z=\bra{1}L\ket{1}+\bra{1}L'\frac1{z-\tilde L}L\ket{1}
.
\ee
The eigenvalue of interest, $\gamma_m$, is positioned near zero and far away, in a relative sense, from other eigenvalues, whereas the corresponding eigenvector is close to $\ket{1}$. Therefore, $\gamma_m$ can be estimated by setting $z=0$ in the denominator of $\frac1{z-\tilde L}$ and taking $\bra{1}L\ket{1}$ and $L\ket{1}$ at second and first order in $\delta_T$, respectively, which gives the result in Eq.\eqref{generalpert}. Since $L^{(1)}\ket{1}$ is orthogonal to $\ket{1}$, we can ignore the distinction between $\tilde L$ and $L$ in the denominator of the second term to simplify the operator calculus below.

We note parenthetically
a direct analogy between the above analysis and the procedure used to compute Green's function of a quantum particle $G(\varepsilon)=1/(\varepsilon-H)$ in terms of its self energy. The latter satisfies Dyson equation, which is an exact relation derived by resumming perturbation series that 
has the same structure as the above formula for the resolvent $R_{11}(z)$. Similar to Dyson equation, which provides a useful tool for developing perturbation theory for a particle which is weakly coupled to other quantum states in the system, we can use the exact form of the resolvent to account for the terms second-order in the off-diagonal part of $L$.

%
%
%
%


Next,
the second-order perturbation result for the eigenvalue $\gamma_m$, given by
the operator expression in Eq.\eqref{generalpert}, needs to be simplified by bringing it to a form that will facilitate the calculations below.
This can be done by making use of the relation in $\eqref{firstevec}$. A convenient way to do it is to multiply and divide the second term by itself and transform it in such a way that the unknown factor $\lambda$ drops out:
\begin{align}\nonumber
& \bra{1}L^{(1)} \frac1{\tilde L^{(0)}
} L^{(1)} \ket{1}=\frac{\bra{1}L^{(1)} \frac1{\tilde L^{(0)}
} L^{(1)} \ket{1}^2}{\bra{1}L^{(1)} \frac1{\tilde L^{(0)}
} L^{(1)} \ket{1}}
\\
&
=\frac{\lambda^2 \bra{1}L^{(1)}\ket{u}^2}{\lambda^2 \bra{u} L^{(0)}\ket{u}}
=\frac{\bra{1}L^{(1)}\ket{u}^2}{\bra{u} L^{(0)}\ket{u}}
.
\end{align}
In the numerator, we used $L^{(1)}\ket{1} =\lambda L^{(0)} \ket{u}$ only when acting on the right whereas in the denominator we used it on the right and left; we also dropped tilde sign on the account of orthogonality of $\ket{1}$ and $L\ket{1}$.
The cancellation of the proportionality constant $\lambda$ between the numerator and denominator 
can also be verified 
by noting that the final expression is invariant under rescaling $\ket{u}$.
%
%
We can write the result above as $\left(L^{(1)}_{1u}\right)^2/L^{(0)}_{uu}$ where we introduced the notation $O_{\eta' \eta} = \bra{\eta'} O \ket{\eta}$.
We therefore have
\begin{equation}
  - \gamma_m^{(2)} = L^{(2)}_{11} - \frac{\left(L^{(1)}_{1u}\right)^2}{L_{uu}}.
  \label{gamma2}
\end{equation}
We note that each of the matrix elements in the above are evaluated at lowest non-vanishing order.  In the next section we are going to evaluate these matrix elements.  We find that the combination does not vanish in general, and that $\gamma_m \propto m^4\log m  \,T^4/T_F^3$ for $T_F/T \gg m^2 \gg 1$.

It is interesting to note that the two terms in Eq.\eqref{gamma2} can be interpreted in terms of the coupling between radial/angular relaxation noted above, wherein the first term accounts for angular relaxation and the second term describes the effects due to radial/angular coupling that partially compensate  that of the angular relaxation. We will see this in more detail below. Here we 
highlight one useful byproduct (and a consistency check) of this analysis:
for $m = \pm 1$ the two terms cancel exactly, giving $\gamma_m^{(2)} = 0$, as expected by momentum conservation.





\section{Matrix element evaluation away from  $\vec q\approx 0$}
\label{sec:matrix_element_evaluation}

The goal of this section will be to compute the matrix elements $L_{uu}$, $L_{1u}$, and $L_{11}$ in \eqref{gamma2}, obtaining $\gamma_m \sim \frac{T^4}{T_F^3}m^4 \log m$.  This is done by expanding the combinations $\sum'_\alpha \eta(\vec{p}_\alpha)$ in the dimensionless energy deviations from the Fermi energy, $\delta_T u_\alpha$ for $\eta(\vec{p}) = 1e^{im\theta_p}$.  These combinations are expanded to first order, so that the quantities $L_{uu}$, $L_{1u}$, and $L_{11}$ are expanded to zeroth, first, and second order in temperature respectively, as in \eqref{gamma2}.

We start with reproducing, for reader's convenience, the expression \eqref{schematicmelt}, with the angular part of the measure $d\nu_{iji'j'}$ expanded:
\begin{align} \label{qelt}
L_{\eta'\eta} &  = \frac{-m_*T^2}{32\pi^3\hbar}\sum_{s_1, s_2} \int d \mu_{iji'j'}
\\
& \times \int \frac{dq}{qv_iv_j\abs{\sin \theta_i \sin \theta_j}}|V|^2\sideset{ }{'}\sum_\alpha \overline{\eta'_\alpha}\sideset{ }{'}\sum_\beta \eta_\beta.
\nonumber
\end{align}
where the integration limits are specified below Eq.\eqref{qomega}.
Anticipating that the dominant contribution arises from the near-head-on collision processes, we begin with a natural starting point: take the above expression and expand around head-on values via the expressions in \eqref{angles}.

The approach to carry out this expansion, developed in this section, will only work sufficiently far away from $q\approx 0$ (namely, for $q\gg 2k_F\delta_T$).
In future sections we  will show that this gives a correct 
answer for scaling with $T$ and $m$ up to an overall constant factor. The analysis of this section 
will also motivate many of the manipulations we do in subsequent sections, where the final, more precise, 
analysis is presented.

\begin{figure}[t]
\includegraphics[width=0.99\columnwidth]{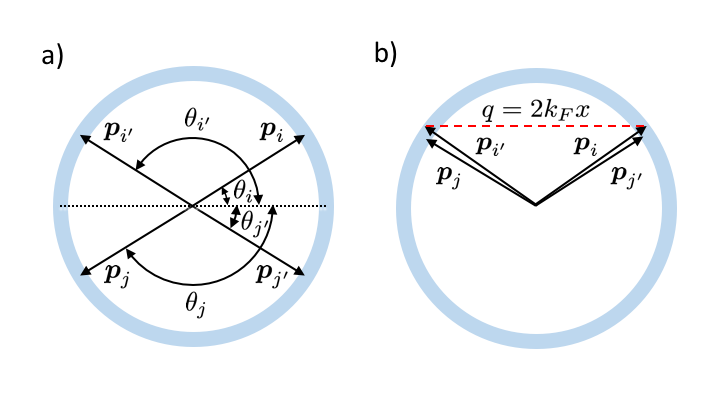}
\caption{
Collision processes analyzed in Sec.\ref{sec:matrix_element_evaluation}.  In panel a), the particles collide nearly head-on whereas in panel b) the particles nearly exchange momenta.  Momenta angles $\theta_\alpha$ are measured with respect to the $\vec q$ axis.
}
 \label{fig4}
\end{figure}

In order to carry out the expansion of the combination $\sum_\alpha' \overline{\eta'_\alpha}\sum_\beta' \eta_\beta$ in $\delta_Tu_\alpha$, it is useful to break it down in terms of cosines and sines.
In doing so, we will continue to use the convention introduced in Sec.\ref{sec:kinematic_constraints}, measuring
all angles $\theta_\alpha$ with respect to $\vec q$.
We decompose $\eta(\vec{p}_\alpha) = g(p_\alpha)e^{im \theta_\alpha} = g(p_\alpha)(\cos m\theta_\alpha + i \sin m\theta_\alpha)$ where $g(p_\alpha)$ for our purposes will be $1$ or $u_\alpha$.  The sum over $s_1$ and $s_2$ means that any terms odd in $\theta$ 
yield zero and so we only have to consider the terms with two cosine factors or two sine factors:
\begin{equation}
  \begin{aligned}
   & \sum_{s_1,s_2} \sideset{ }{'}\sum_\alpha  \overline{\eta}'_\alpha \sideset{ }{'}\sum_\beta \eta_\beta  \\
   &= \sum_{s_1,s_2}  \bigg(  \sideset{ }{'}\sum_\alpha g_\alpha'\cos m \theta_\alpha \sideset{ }{'}\sum_\beta g_\beta \cos m \theta_\beta
  \\
& \qquad  + \sideset{ }{'}\sum_\alpha g_\alpha'\sin m \theta_\alpha \sideset{ }{'}\sum_\beta g_\beta \sin m \theta_\beta \bigg).
  \end{aligned}
  \label{sincosdecomp}
\end{equation}
We now proceed to expand the cosine terms, for the time being ignoring the
sine terms.  The contribution of these terms is dominated by the small-$q$ processes, for which the approach used in this section is not valid. The sine terms will be analyzed below in two different ways, by mapping on the cosine terms in Sec.\ref{sec:duality} and then in a more direct way in Sec.\ref{sec:full_calculation}.

To carry out the expansion we define 
a frequency parameter which is ``dual'' to $w$:
\begin{equation}
  \tilde{w} = \frac{\tilde{\omega}}{T} = \frac{\varepsilon_i - \varepsilon_{j'}}{T} = \frac{\varepsilon_{i'} - \varepsilon_j}{T}
  \label{dualwdefn}
\end{equation}
(the duality nature of the relation between the quantities $w$ and $\tilde w$ will become clear in Sec.\ref{sec:duality}).
We can now expand the right hand sides of the relations given in Eq.\eqref{angles} to first order in $w$ and $\tilde w$ as
\begin{equation}
  \begin{aligned}
    \cos \theta_i    & \approx +x + \frac{\delta_T}{4}\left( yw - x\tilde{w} \right), \\
    \cos \theta_j    & \approx -x + \frac{\delta_T}{4}\left( y w - x\tilde{w} \right), \\
    \cos \theta_{i'} & \approx -x + \frac{\delta_T}{4}\left( y w + x\tilde{w} \right), \\
    \cos \theta_{j'} & \approx +x + \frac{\delta_T}{4}\left( y w + x\tilde{w} \right)
  \end{aligned}
  \label{qangleexpand}
\end{equation}
where we defined $y=\frac{1}{x} - x$ and $x$ is defined in \eqref{xandtheta}.
Using the Chebyshev polynomials of the first kind, $\cos m\theta_\alpha = T_m(\cos \theta_\alpha)$, we have
\begin{equation}
\begin{aligned}
 \cos m \theta_i   & \approx  + T_m(x) + \frac{\delta_T}{4}T'_m(x)\left( y w - x\tilde{w} \right), \\
    \cos m \theta_j   & \approx  - T_m(x) + \frac{\delta_T}{4}T'_m(x)\left( y w - x\tilde{w} \right), \\
    \cos m \theta_{i'}& \approx  - T_m(x) + \frac{\delta_T}{4}T'_m(x)\left( y w + x\tilde{w} \right), \\
    \cos m \theta_{j'}& \approx  + T_m(x) + \frac{\delta_T}{4}T'_m(x)\left( y w + x\tilde{w} \right).
  \end{aligned}
  \label{qchebyTexpand}
\end{equation}
It is important to note that this expansion is only valid for $m^2\delta_T \ll 1$ since otherwise higher derivatives of $T_m(x)$ would become equally important close to $x=1$.

Carrying out the expansion for 
the state $\ket{u}:=\frac{u}2 e^{im\theta}$, 
{yields a nonvanishing zeroth order value \begin{equation}
  \begin{aligned}
  \sideset{}{'}\sum_\alpha \frac{u_\alpha}{2} \cos m \theta_\alpha & \approx \frac{1}{2}(u_i - u_j + u_{i'} - u_{j'})T_m(x) \\
  & = \tilde{w} T_m(x)
.
  \end{aligned}
  \label{ustatechebyTq}
\end{equation}
For the state $\ket{1}:=e^{im\theta}$, in contrast, we have cancellation to zeroth order, as expected.  We 
are therefore left with the first order contribution
\begin{equation}
  \sideset{}{'}\sum_\alpha \cos m \theta_\alpha \approx -\delta_T\tilde{w}xT'_m(x).
  \label{1statechebyTq}
\end{equation}
To obtain the quantities $L_{uu}$, $L_{1u}$, and $L_{11}$, these expressions must be substituted in Eq.\eqref{qelt} and integrated over momentum transfer ($\int dq...$),
and then over energies ($\int d\mu_{iji'j'}...$). 

We first discuss the strategy for $q$ integration, arguing that the result is dominated by $q\approx 2k_F$.
Focusing on $x=\frac{q}{2k_F}\approx 1$, and simplifying the measure accordingly, will help us to carry out the integration in a closed form with logarithmic accuracy.
We first note that the naive simplification of the denominator in the $q$ integration in \eqref{qelt} gives
\be
\frac{dq}{qv_iv_j\abs{\sin \theta_i \sin \theta_j}} \approx \frac{dx}{
v_F^2x(1-x^2)}
,
\ee
featuring divergences as $x \to 0$ and $x \to 1$.  The former divergence is not a problem since both \eqref{ustatechebyTq} and \eqref{1statechebyTq} vanish as $x \to 0$.  The limit $x\to1$ must be treated with some more care, however.  We note parenthetically that the convergence at $x=0$ is a convenient feature of the $\cos m \theta_\alpha$ terms, on which we focus in this section. The $\sin m \theta_\alpha$ terms, to the contrary, lead to quantities that are finite as $x \to 0$ and go to zero as $x \to 1$.  The $x\to 0$ limit is problematic as the terms $\delta_Tw\lp \frac{1}{x} -x \rp$ are no longer small for sufficiently small $x$, and the perturbation theory breaks down.  We will remedy this problem in subsequent sections.

To perform the $q$ integration in \eqref{qelt} we must cure the divergence from $x\to1$.  To do this, we include the first order terms in $\delta_T$ for $\sin \theta_i$ and $\sin \theta_j$.  The $\delta_T w \lp \frac{1}{x}-x \rp $ terms vanish as $x \to 1$ and so can be ignored.  Noting that the bounds in the $q$ integration in Eq.\eqref{qelt} are such that we integrate until $\abs{\sin \theta_i \sin \theta_j} = 0$, we obtain
\begin{widetext}
\be
\begin{aligned}
& \int \frac{dx}{v_iv_jx\abs{\sin \theta_i \sin \theta_j}} \cdots  = \int\limits_0^{1-a^2/2} \frac{dx}{v_F^2x\sqrt{1-\lp x-\frac{a^2x}{2}\rp^2}\sqrt{1-\lp x+\frac{a^2x}{2}\rp^2}} \cdots  = \int\limits_0^{1-a^2/2} \frac{dx}{v_F^2x(1-x^2)} \cdots, \\
\end{aligned}
\label{jacobian}
\ee
\end{widetext}
where
\begin{equation}
  a = \sqrt{\delta_T \abs{\tilde{w}}/2}
  \label{adef}
\end{equation}
is the minimum value of $\theta = \arccos x$, the quantity defined in \eqref{xandtheta}.
The bounds in \eqref{jacobian} are obtained to first order in $a^2$ which is sufficient to cure the divergence with logarithmic accuracy and ensure that further corrections would only give higher order corrections to the entire integral.  The resulting integrand does not follow pointwise from the first or second integrands but gives the same answer at log order after integration because the rest of the integrand is approximately constant in the region where $1-x \sim a^2/2$.

Putting everything together, we change variables from $q$ to $\theta$
and write the quantities $L_{\eta'\eta}$ as
\begin{equation}
  L_{\eta'\eta} = \frac{- m_* T^2}{32\pi^3\hbar v_F^2}\int d\mu_{iji'j'} \tilde{w}^2  \sum_{s_1, s_2}J_{\eta\eta'}
  \label{finaleltq}
\end{equation}
where we introduced the quantities
\begin{widetext}
\begin{equation}
  \begin{aligned}
    J_{uu} & = \int_0^{1-\frac{a^2}{2}} \frac{dx}{x(1-x^2)}\abs{V}^2 T_m(x)^2 = \int_a^{\pi/2} \frac{d\theta}{\cos \theta \sin \theta }\abs{V}^2\cos^2{m\theta} = \int_a^{\pi/4} \frac{d\theta}{\cos \theta \sin \theta }\abs{V}^2,
  \end{aligned}
  \label{Jeltsq1}
\end{equation}
\begin{equation}
  \begin{aligned}
    J_{1u} & = -\delta_T m \int_0^{1-\frac{a^2}{2}}\frac{dx}{x(1-x^2)}\abs{V}^2 T_m(x)T'_m(x) = -\delta_T m\int_a^{\pi/2} \frac{d\theta}{\cos \theta \sin \theta}\abs{V}^2\cot \theta \cos m\theta \sin m \theta
\\
&= -\delta_T m\int_a^{\pi/4} \frac{d\theta}{\cos^2 \theta \sin^2 \theta}\abs{V}^2  \cos m\theta \sin m \theta,
  \end{aligned}
  \label{Jeltsq2}
\end{equation}
\begin{equation}
  \begin{aligned}
    J_{11} & = \delta_T^2 \int_0^{1-\frac{a^2}{2}} \frac{dx}{x(1-x^2)}\abs{V}^2 T'_m(x)^2= \delta_T^2 m^2 \int_a^{\pi/2} \frac{d\theta}{\cos\theta \sin\theta}\abs{V}^2\cot^2\theta \sin^2 m\theta \\
    & = \delta_T^2 m^2 \int_a^{\pi/4} \frac{d\theta}{\cos\theta \sin\theta}\abs{V}^2\left(\cot^2\theta \sin^2 m\theta \right.
 \left.
+ \tan^2 \theta \cos^2 m \theta\right).
  \end{aligned}
  \label{Jeltsq3}
\end{equation}
\end{widetext}
Here we simplified the dependence on $\theta$ under the integrals by  symmetrizing it with respect to $\theta\to\pi/2-\theta$.

Now everything is in place to estimate $\gamma_m$ values. First, as a quick validity check, we can plug in $m=\pm1$ and find that $J_{11} = \delta_T J_{1u} = \delta_T^2 J_{uu}$, and so $\gamma_{\pm1} =0$ by \eqref{gamma2}.  For other values of $m$ we will generically
obtain nonzero values for $\gamma_m$.  These can be estimated for $m \gg 1$ by asymptotically expanding the above integrals.  In particular, for each integral we have a logarithmic divergence that is cut off at one end by $a$.  For $J_{uu}$ the integrand is order $1$ over an order $1$ range of $\theta$, whereas for $J_{1u}$ and $J_{uu}$ the integrands are order $m^2$ and $m^4$ respectively for a range of $\theta$ on the order of $1/m$.
While the matrix element $\abs{V}^2$ 
in general depends on $s_1, s_2$, in the limit $x\to1$ it is $\abs{V_*}^2$.
We therefore obtain
\begin{equation}
  \begin{aligned}
    \sum_{s_1, s_2}J_{uu} & = 4\abs{V_*}^2\log\frac{1}{a} + \cdots, \\
    \sum_{s_1, s_2}J_{1u} & = -4\delta_T m^2\abs{V_*}^2\log \frac{1}{ma} + \cdots, \\
    \sum_{s_1, s_2}J_{11} & = 4\delta_T^2 m^4\abs{V_*}^2\log \frac{1}{ma} + \cdots.
  \end{aligned}
  \label{asymptoticq}
\end{equation}
Splitting $\log\frac{1}{ma} = \log a^{-1} - \log m$ and asymptotically expanding with $a^{-1} \gg m$, we obtain from \eqref{gamma2} that the terms proportional to $\int d\mu_{iji'j'}\tilde{w}^2\log a^{-1}$ vanish.  The $\log m$ terms do not cancel out however.  Noting that
\begin{equation}\label{eq:8pi4/15}
  \int d\mu_{iji'j'} \tilde{w}^2 = \frac{8\pi^4}{15},
\end{equation}
[see \hyperref[appendix:integrals]{Appendix}] we arrive at
\begin{equation}
\begin{aligned}
  \gamma_m & = \frac{\pi m_*^2 \abs{V_*}^2k_B}{15\hbar^5}  \frac{T^2}{T_F} \delta_T^2 m^4 \log m \\
  & = \frac{\pi m_*^2 \abs{V_*}^2k_B}{15\hbar^5}\frac{T^4}{T_F^3}m^4 \log m,
  \label{gammaq}
\end{aligned}
\end{equation}
where we converted all dimensionful quantities to factors of $\hbar$, $k_B$, mass $m_*$ and degeneracy temperature $T_F$.  This expression
scales with respect to $T$ and $m$ as anticipated above.

Before closing this section we mention two technical issues with the above analysis that still need to be addressed. One is that we ignored the $\sin m \theta_\alpha$ terms. The other is that the analysis appears to break 
down when $x$ becomes of order $\delta_T$. These shortcomings will be resolved as follows. In Sec.\ref{sec:duality} we will show that the $\sin m \theta_\alpha$ terms can be mapped onto the $\cos m \theta_\alpha$ terms.  We will also show that the $\cos m \theta_\alpha$ terms still receive no contribution from 
the $x \sim \delta_T$ processes once the latter are treated properly.  This implies that the only correction to the above result is a factor of $2$.  In Sec.\ref{sec:full_calculation} we will also redo  the whole calculation in a different, more logical way, 
with these complications taken into account from the start.

\section{Discussion of superdiffusive result and radial corrections}
\label{sec:discuss_superdiffusion}

Before we move onto patching up the technical issues in the above analysis, we 
take a moment to discuss the physical picture that emerged from the 
our discussion. One interesting aspect is the relation between the $m^4$ dependence and angular diffusion. Another is related to roles of the $\ket{1}$ and $\ket{u}$ states, which account for radial relaxation and for the interplay between the latter and angular relaxation.

To understand the relation between the $m^4$ dependence and angular diffusion, we recall that we found that the odd-$m$ relaxation is dominated by scattering processes representing perturbations of forward collisions and head-on collisions. This is an interesting situation, since neither of these collision types, taken per se, have any impact on odd-$m$ harmonics, and yet the near-forward and near-head-on collisions dominate the relaxation dynamics. The perturbations about forward and head-on collisions 
are described by small angular steps on a circle resulting from each scattering event, which calls very naturally for an angular diffusion interpretation. As discussed below, such a diffusion picture can indeed be constructed, however with two caveats.

One caveat is that, since only the odd-$m$ harmonics of the distribution are involved in the dynamics, we must identify ingoing (outgoing) particle states with angle $\theta$ with an outgoing (ingoing) particle with angle $\theta + \pi$, respectively. Namely, the configuration space for this diffusion process is a circle with the points $\theta$ and $\theta+\pi$ glued together, which is still a circle, albeit of a twice smaller  circumference. This allows us to think about near-head-on collision processes in terms of small angular steps in configuration space.

Another caveat is related with the angular step size dependence on  momentum transfer $q=2k_F x$. For most values of $x$ we obtain from the above analysis a step size $\Delta \theta \sim \delta_T$ and a factor of $m$ from $T_m'(x)$.
Angular diffusion with the diffusion coefficient $D_\theta=\frac{T^2}{\varepsilon_F}\delta_T^2$ would then predict a relaxation rate
$\gamma_m \sim D_\theta m^2=\frac{T^2}{\varepsilon_F}m^2\delta_T^2$ for these collisions. However, we find interesting behavior as $x \to 1$ with the step size becoming anomalously large. Indeed, as $x \to 1$ the step size $\Delta \theta$ is no longer of order $\delta_T$ but instead it gets gradually enhanced to $\sqrt{\delta_T}$ because of the flatness of $\cos \theta_\alpha$ for $x$ close to $1$.
It is not enough, however, to simply replace the step size in a one-particle picture. Instead, we must account for the angular diffusion changing character from one-particle random walk to a correlated two-particle dynamics.


The origin of this correlated behavior can be seen as follows. We recall that the rate $\gamma_m$ in our analysis was found to be dominated by the contribution of $x\to 1$, where several interesting things happen.
In particular, $T_m'(x)$ becomes of order $m^2$ which leads to $\gamma_m \sim \frac{T^2}{\varepsilon_F}m^4\delta_T^2$ 
a behavior distinct from that expected from standard diffusion.  This arises because of two effects.
One is the enhancement of the angular step  from $\delta_T$ to $\sqrt{\delta_T}$, mentioned above. The value $\sqrt{\delta_T}$ is considerably greater than the width $\delta_T$ of thermally smeared Fermi surface. 
The large angular step size 
$\sqrt{\delta_T}$ comes with 
a second effect --- {\it nontrivial two-particle correlations.} 
Indeed, by momentum conservation in the direction perpendicular to $q$ we must have $\theta_i - \theta_j + \theta_{i'} - \theta_{j'} \to 0$ as $x \to 1$.  In other words, even though the stepsize is increased from $\delta_T$ to $\Delta\theta=\sqrt{\delta_T}$, momentum conservation forces a correlation between the two angular steps such that they cancel each other out to lowest order,
\be
\Delta\theta_{i'}=-\Delta\theta_{j'}
.
\label{eq:angular_correllations}
\ee
Therefore, a ``one-particle diffusion constant", naively estimated as $D=\frac{T^2}{T_F}(\Delta\theta)^2\sim \frac{T^3}{T_F^2}$, and the associated rates $\gamma_m=Dm^2$ do not provide a correct answer. Instead,
we have a ``one-particle diffusion constant" of zero  in the $x \to 1$ limit, and
the correct procedure must account for the correlations, Eq.\eqref{eq:angular_correllations}. This is precisely what the expansion carried out in Secs.\ref{sec:perturbation_theory} and \ref{sec:matrix_element_evaluation} is doing, arriving at
the fourth order term with an enhanced stepsize contributing in the limit $x \to 1$, such that $\gamma_m \sim \frac{T^2}{\varepsilon_F}m^4(\Delta \theta)^4 \sim \frac{T^2}{\varepsilon_F}m^4\delta_T^2$.  We call this fourth order but enhanced stepsize diffusion ``superdiffusion'', since its net effect is to enhance the $m^2$ scaling to $m^4$.  The use of Chebyshev polynomials automatically combines the effects of increasing stepsize and decreasing diffusion constant such that all we see is a smooth transition from ordinary diffusion to superdiffusion as $x$ approaches $1$.

The above discussion in terms of angular diffusion and superdiffusion explains the $m$ dependence and temperature dependence fairly well, but it is important to emphasize that it is not the full story.  Most obviously, 
it explains neither the logarithmic factors nor momentum conservation.
Another, perhaps related, point is that a full understanding requires understanding radial relaxation which accompanies angular relaxation. 
Radial relaxation is included in the second-order perturbation theory through transitions between the $\ket{1}$ and the first-order eigenvector correction $\ket{u}$, 
Eq.\eqref{eq:eigenvector_correction_eta1}, generating the negative term in Eq.\eqref{gamma2}. The fine balance between the two terms, negative and positive, is essential in our analysis.

From a qualitative standpoint, 
one can say that the constraints of momentum and energy conservation mandate that {\it every angular step comes with a radial step}.  In this sense, momentum conservation demands the inclusion of transitions between the $\ket{1}$ and $\ket{u}$ modes.  For $m=1$ this inclusion repairs momentum conservation through a perfect cancellation between the negative and positive terms in Eq.\eqref{gamma2}. 
The interplay between angular and radial dynamics is also important for higher $m$, 
as it impacts
numerical and logarithmic prefactors for the rates $\gamma_m$.
One might worry that the superdiffusive behavior would be canceled out due to these corrections, as the $\log \frac{1}{a}$ prefactor of it is, but instead we find that it survives with a residual coefficient of $\log m$.


\section{Extending the analysis to small $\vec q$: duality transformation and reflections}
\label{sec:duality}


It might seem that the analysis done so far is very incomplete, since Sec.\ref{sec:matrix_element_evaluation}
only handles some contributions (cosine terms rather than sine terms, large $q$ rather than any $q$). The goal of this section is to vindicate the analysis of Sec.\ref{sec:matrix_element_evaluation}. This is done by invoking a suitably defined duality transformation and reflection symmetry in order to map the contributions that were ignored in Sec.\ref{sec:matrix_element_evaluation} onto to the ones that were analyzed.
We will see that the evaluation of matrix elements carried out in Sec.\ref{sec:matrix_element_evaluation} gives the correct final answer up to a proportionality constant (factor of two). 
At this stage, however, this is far from obvious.
Most obviously, the $\sin m \theta_\alpha$ terms were ignored. Even for cosine terms, which we did consider in Sec.\ref{sec:matrix_element_evaluation}, small $q$ processes 
were not treated with care, as the expansions in \eqref{qchebyTexpand} may break down as $x \to 0$.
Since the measure $dq/q$ is scale invariant, these processes have just as large of a phase space as the processes with $q$ on the order of $k_F$.

To begin cataloging the important processes missed by
the expansion in \eqref{qchebyTexpand} 
we consider what happens when particle $i'$ is switched with particle $j'$.  For $s_1 = -s_2$, this exchange takes an almost head-on process to another almost head-on processes, both of which are integrated over as $q$ varies from $0$ to $2k_F$.  However, for $s_1 = s_2$ we always have an exchange process with $\vec{p}_i \approx \vec{p}_{j'}$ and $\vec{p}_{i'} \approx \vec{p}_j$, as shown in Fig. \ref{fig2} and Fig. \ref{fig3}.  Exchanging $i'$ and $j'$, we obtain a forward scattering processes with $\vec{p}_i \approx \vec{p}_{i'}$ and $\vec{p}_j \approx \vec{p}_{j'}$. This process is unjustifiably missed in the expansion in $\delta_T w/x$ in \eqref{qchebyTexpand};
Indeed, the forward scattering process has $q \approx 0$, and in particular $x \sim \delta_T$.  This is where we expect our
expansion 
to break down, and so it is not a surprise that this process was missed.
We also know
the missed forward process must give the same contribution as an exchange process since it differs only by exchanging identical particles.


In this section we 
account for  these processes. 
as well as all other relevant ones, by mapping them onto ones we have considered in Sec.\ref{sec:matrix_element_evaluation}.  We also include the $\sin m \theta$ terms by relating them to the $\cos m \theta$ terms that we have already considered.  The end result is that the $\sin m \theta$ terms with the new processes give the same contribution as the $\cos m \theta$ terms with the large $q$ collisions.  The small-$q$ processes give negligible contributions to the $\cos m \theta$ terms, and 
the same is true for the contribution of large-$q$ processes to the $\sin m \theta$ terms.  The end result, as we will see, is just an overall factor of $2$.

In our discussion, we will make extensive use of the fact that the tips of the four momentum vectors involved in a two-body scattering process coincide with 
four corners of a rectangle, with $\vec{p_i}$ positioned across the diagonal from $\vec{p}_j$ and $\vec{p}_{i'}$ positioned across the diagonal from $\vec{p}_{j'}$.  This rectangle property can be interpreted in terms of a geometric {\it duality transformation}.

The rectangular arrangement is a simple consequence
of kinematics of two-body collisions combined with parabolic dispersion $\varepsilon=p^2/2m-\varepsilon_F$. 
In this case the ``dual'' momentum transfer defined by exchanging particles $i'$ and $j'$,
\begin{equation}
  \tilde{\vec{q}} = \vec{p}_i - \vec{p}_{j'} = \vec{p}_{i'} - \vec{p}_j
  \label{tildeq}
\end{equation}
is always perpendicular to the momentum transfer $\vec q=\vec p_i-\vec p_{i'}$ used above:
\be\label{eq: rectangle}
\tilde {\vec q}\perp \vec q
.
\ee
Two of the sides of the rectangle are the momentum transfer $\vec{q}$ and the other two sides are the ``dual'' momentum transfer $\tilde{\vec{q}}$.
One can see that the momenta form a rectangle by boosting to the center of mass frame.
In this frame, we have a perfect head-on collision, $\vec{p}_i = - \vec{p}_j$, $\vec{p}_{i'} = -\vec{p}_{j'}$, and
therefore the tips of the momenta form a rectangle.

The rectangle property can be seen e.g. in Fig.\ref{fig3} and Fig.\ref{fig4} above; however, the rectangles formed by $\vec q$ and $\tilde{\vec q}$ are not shown in these figures to avoid overcrowding. The rectangle property will be central to our discussion in the next section; it is exhibited explicitly e.g. in Fig.\ref{fig5}.

Having introduced $\tilde q$ we can define a new ``dual'' coordinate system in which instead of $s_1$ and  $s_2$ the scattering particles configurations are labeled by $\tilde s_1$ and  $\tilde s_2$. The variables $\tilde{s}_1$ and $\tilde{s}_2$ did not appear explicitly in Sec.\ref{sec:matrix_element_evaluation} 
because we privileged $q$ over $\tilde{q}$. Nevertheless they are in principle accounted for in the detailed behavior at $q\to 0$ of \eqref{angles} (this behavior was 
ignored earlier when we performed perturbation theory).  We note that the integration measures $d\nu_{iji'j'}$ and $d\mu_{iji'j'}$ do not change under reflections corresponding to reversing signs of 
 $s_1$ or $s_2$, and so they do not change under 
reversing signs of the dual 
quantities $\tilde{s}_1$ and $\tilde{s}_2$ either.

The name ``duality" is used here because of an interpretation of  the rectangular arrangement in terms of interchanging the particle outgoing states $i'$ and $j'$ while leaving $i$ and $j$ intact. Switching $i'$ and $j'$ makes no difference from the kinematic constraints point of view; however it is equivalent, geometrically, to switching the long and short sides of the rectangle.  Note that $w$ and $\tilde{w}$ are related by duality as well. We also note that the form of the matrix element \eqref{eq:V=U-U} is antisymmetric under duality, as required by Fermi statistics, and so $\abs{V}^2$ is invariant.



We note parenthetically that, while the rectangular geometry of collisions with parabolic dispersion appears to be used in a crucial way here,
in fact we have chosen to work with parabolic dispersion only for convenience.  Since 
all the
momenta are only expanded linearly 
about the Fermi level, our analysis must be insensitive to the type of dispersion and can only depend on the slope of the dispersion relation at the Fermi level (i.e. the effective mass $m_* = p/v$). We therefore believe that the Galilean symmetry associated with the parabolic dispersion is not essential for the conclusions of our analysis.



The geometric observations based on duality make it obvious that the $\sin m \theta_\alpha$ terms do not generate anything different from $\cos m \theta_\alpha$ terms. Indeed, if
\begin{equation}
  \tilde{\theta}_\alpha = \theta_\alpha-\pi/2
  \label{tildephi}
\end{equation}
is the angle between $\vec{p}_\alpha$ and $\tilde{\vec{q}}$, then because $m$ is odd we have $\sin m \theta_\alpha = (-1)^{\frac{m-1}{2}}\cos m \tilde{\theta}_\alpha$.
Thus, we can just switch $i'$ and $j'$ to turn all $\sin m \theta_\alpha$ terms to $\cos m \theta_\alpha$ terms (the sign in front cancels out in \eqref{sincosdecomp}).  Technically, we could also have $\tilde \theta_\alpha = \theta_\alpha + \pi/2$ if $\tilde{\vec q}$ points in the opposite direction, but this only results in a different sign canceling.
We are therefore justified  in only considering the $\cos m \theta_\alpha$ terms and multiplying the integral by $2$.

We have discussed how the analysis in Sec.\ref{sec:matrix_element_evaluation} misses processes that have small momentum transfer $q$, and how some of these processes can be analyzed if we use the dual momentum transfer $\tilde{q}$ to label collisions instead.  We now discuss how all small $q$ processes can be included and why they do not contribute to the $\cos m \theta_\alpha$ terms.
In the last section, $s_1 = s_2$ led to very thin and long rectangles where $q$ was large and $\tilde{q}$ was small.  These were not an issue since our expansions were valid as long as $q$ was large.  But since we privileged $q$ over $\tilde{q}$, we did not see the collisions where $q$ was small but $\tilde{q}$ was large, where the dual analogues to $s_1$ and $s_2$, introduced above as $\tilde{s}_1$ and $\tilde{s}_2$, are equal.  This leads to the notion that we should use a $\tilde{q}$ coordinate system for these collisions instead.  However, the issues are slightly more subtle than this, since there are also collisions where both $q$ and $\tilde{q}$ are small that are missed in \emph{both} coordinate systems (where $s_1 = s_2$ and $\tilde{s}_1 = \tilde{s}_2$).  These are nearly collinear processes where the entire collision rectangle is inside the Fermi surface broadening. We will use reflection symmetry and energy conservation to show that these collisions must have vanishing contribution.

To resolve this issue, 
in addition to applying the duality we also employ mirror reflections across $\vec{q}$ or $\tilde{\vec{q}}$.  In particular, for any collision with both $q$ and $\tilde{q}$ small, we can temporarily 
reverse the sign of one of $s_1, s_2, \tilde{s}_1, \tilde{s}_2$ so that one of $q$ or $\tilde{q}$ 
becomes large.  We can then use the Chebyshev expansions to compute the angles and then
reflect back to the original collision geometry by adding in negative signs where necessary.  This way of treating the collisions allows us to show that none of the missed collisions contribute appreciably to the $\cos m \theta$ terms.

\begin{figure}[t]
\includegraphics[width=0.99\columnwidth]{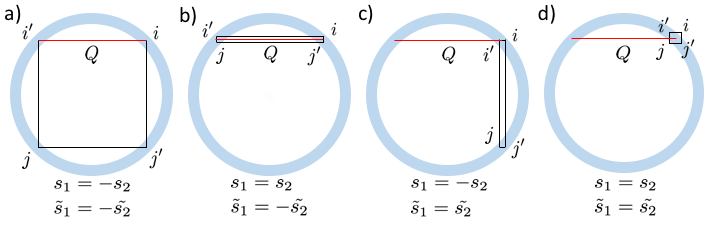}
\caption{Illustration of the labels $Q$; $s_1$, $s_2$; and $\tilde{s}_1$,  $\tilde{s}_2$ defined in and nearby Eqs.\eqref{Qfromreflections},\eqref{Qdefn}, Eq.\eqref{s1s2}, and Eq.\eqref{dualreflections} respectively.  We use rectangles to illustrate the collision as discussed around Eq.\eqref{tildeq}.
The momentum vectors end at the corners of the rectangles and the horizontal and vertical side lengths correspond to $q$ and $\tilde{q}$ respectively.  Panels $\textbf{a}$ and $\textbf{b}$ label head-on and exchange collisions that were analyzed in previous sections and have large $q$.  Panels $\textbf{c}$ and $\textbf{d}$ have small values of $q$ and were missed in the previous analysis, but have large values of $Q$ (the same as \textbf{a} and \textbf{b} and can thus now be included.
}
 \label{fig5}
\end{figure}

We now describe the procedure outlined in the above paragraph in more detail and show why the missed collisions have vanishing contributions to the $\cos m \theta_\alpha$ terms.  Any missed collision in the previous section's perturbation theory can be obtained by reflecting one of the pairs $\vec{p}_i, \vec{p}_{j'}$ or $\vec{p}_j, \vec{p}_{i'}$ across the line defined by $\tilde{\vec q}$ such that after the reflection $\tilde{s}_1 = -\tilde{s}_2$.
After reflection, the collision will then have $\vec p_i$ and $\vec p_{i'}$ on different sides of the $\tilde{\vec{q}}$ axis.  It will then have $q \sim k_F$ unless $\theta_\alpha \approx \pi/2$.
In the former case we can apply the previous sections perturbation theory.
In the latter case, momentum conservation along the $\vec{q}$ axis shows that $\sideset{ }{'}\sum_\alpha \cos m \theta_\alpha \approx \pm m(\tilde{\theta}_i - \tilde{\theta}_j + \tilde{\theta}_{i'} - \tilde{\theta}_{j'}) \approx 0$, and so we can assume the former case and apply perturbation theory on the reflected collision to measure the angles.
Such a reflection 
reverses the sign of $\cos m \theta_\alpha$ if $\alpha$ is one of the two particles with reflected momenta.  However, both of the possible 
sign reversals  lead to the combination $\pm(u_i + u_j - u_{i'} - u_{j'})$ appearing in \eqref{ustatechebyTq} and \eqref{1statechebyTq} and hence the contribution vanishes by energy conservation.  Any of the missed collisions therefore do not appreciably contribute to the $\cos m \theta_\alpha$ terms and we are done.

We note that the matrix element $\abs{V}^2$ depends on $s_1,s_2, \tilde{s}_1,\tilde{s_2}$ as a (symmetric) function of $\vec q$ and $\tilde{\vec q}$, see \eqref{eq:V=U-U}.  The matrix element that contributes at leading order is again $\abs{V_*}^2$ since for $\tilde{s}_1 = \tilde{s_2}$ the $\cos m \theta_\alpha$ terms vanish by energy conservation.

While the above analysis, based on duality and reflections, does fully account for the $q \approx 0$ collisions by mapping them onto what we have already done for collisions away from $q\approx 0$, for completeness we also provide a full derivation that incorporates these ideas from the start. This approach, which will be discussed in detail in the next section, is based on the following idea. To account for the $q \approx 0$ collisions in the above paragraphs, we effectively used a different labeling of collisions that will be made explicit now and in the next section.  Instead of labeling collisions with their (small) momentum transfer $q$ and computing angles in these coordinates, we instead 
reversed signs of $\tilde{s}_1$ and $\tilde{s}_2$ to obtain a different collision with a large momentum transfer, which we will now denote as $Q$ such that \begin{equation}
  Q = \max_{\tilde{s}_1, \tilde{s}_2} \, q.
  \label{Qfromreflections}
\end{equation}
For example, in Fig.\ref{fig5} panels c and d would be mapped onto panels a and b, respectively.  The collisions in all the panels are therefore labeled by $Q$, and angles are initially measured using the collisions with momentum transfer $Q$ (a and b).  The labels $\tilde{s}_1$ and $\tilde{s}_2$ then specify how to recover the angles of the original collision by reflecting from the collision with momentum transfer $Q$.  In particular, for $\tilde{s}_{1} = -\tilde{s}_2$ we have $q = Q$ and the analysis is unchanged.  For $\tilde{s}_1 = \tilde{s}_2$ (panels c or d), we label the collision with $Q$ which is order $k_F$, unlike $q$ which is small, and use the angles measured from the collision with momentum transfer $Q$ (panels a or b) together with reflections to obtain the angles for the collision of interest.


\section{Full calculation}
\label{sec:full_calculation}

We now present a full version of the calculation, where the ideas in the previous section are merged with the calculation details rather than used to repair them afterwards.  As in Sec.\ref{sec:duality}, we use duality and reflections to parameterize collisions in such a way that expansion of Chebyshev polynomials does not break down. To do this, we will need to work towards a representation where all the reflections $s_1, s_2, \tilde{s}_1, \tilde{s}_2$ are manifest.  We work with the collision integral where the unsplit energy delta function is reinstated but the momenta $\vec{p}_{i'}$ and $\vec{p}_{j'}$ have been integrated over by splitting the momentum delta function.  We also don't work with energies normalized by temperature for the moment, opting to reinstate the power counting once we
have arrived at a form of the integral where we can apply perturbation theory.  We have
\begin{align}\nonumber
 & L_{\eta' \eta} = \frac{- m_*\beta}{32 \pi^3 \hbar}\int d\varepsilon_i d\varepsilon_j  d \theta_i d \theta_j f_if_j(1-f_{i'})(1-f_{j'}) \\
& \times \int qdq \delta\left( \sideset{ }{'}\sum_\alpha \varepsilon_\alpha \right)\abs{V}^2 \sideset{}{'}\sum_\alpha \overline{\eta'}_\alpha \sideset{}{'}\sum_\alpha \eta_\alpha,
  \label{fullstart}
\end{align}
where rotation symmetry was used to integrate over $\theta_q$ to cancel the $1/2\pi$ in the inner product.  Instead of splitting the energy delta function with $\omega$ and integrating over it with angles, we instead integrate over it with $q$.  In order to ensure the delta function always has a solution, we allow for negative values of $q$ and divide by $2$.  Using the expresssions for the energies $\varepsilon_{i'} = \varepsilon_i + v_iq\cos \theta_i + \frac{q^2}{2m_*}$ and $\varepsilon_{j'} = \varepsilon_j - v_jq\cos \theta_j + \frac{q^2}{2m_*}$, we obtain the simple result \begin{equation}
  \frac{1}{2}\int_{-\infty}^{\infty} qdq \delta(\varepsilon_i + \varepsilon_j - \varepsilon_{i'} - \varepsilon_{j'}) = m_*.
  \label{parbandcanc}
\end{equation}
Plugging this into the above we have
\begin{align}\nonumber
 & L_{\eta' \eta} = \frac{- m_*^2\beta}{32\pi^3\hbar}\int d\varepsilon_i d\varepsilon_j  d \theta_i d \theta_j f_if_j(1-f_{i'})
\\
& \times (1-f_{j'})\abs{V}^2 \sideset{}{'}\sum_\alpha \overline{\eta'}_\alpha \sideset{}{'}\sum_\alpha \eta_\alpha.
  \label{anglecoords}
\end{align}
This is our starting point for a change of variables to an expression similar to \eqref{qelt}, though with $\tilde{s}_1$ and $\tilde{s}_2$ included and no breakdown of perturbation theory.  In particular, if we use the expressions \eqref{angles} to define a change of variables from $\theta_i$ and $\theta_j$ to $q$ and $\omega$, we obtain \eqref{qelt} with the collisions that have, for example, $\theta_i \in (\pi/2,3\pi/2)$ non-analyzable in perturbation theory because they have small $q$.  We instead come up with a change of variables with an analogue to $q$, denoted as $Q$, such that $Q$ is large for these collisions too.  In particular, we use the following augmentation of $\eqref{qelt}$:
\begin{equation}
  \begin{aligned}
    \abs{\cos \theta_i} = \frac{Q}{2p_i} + \frac{\omega}{v_iQ}, \\
    \abs{\cos \theta_j} = \frac{Q}{2p_j} - \frac{\omega}{v_jQ}.
  \end{aligned}
  \label{Qdefn}
\end{equation}
Note that we can now have $\theta_i \in (\pi/2,3\pi/2)$ without $Q$ small since the absolute value sign enables both signs of $\cos \theta_i$.  Hence, we have the additional labels \begin{equation}
  \begin{aligned}
    \tilde{s}_1 = \sgn(\cos \theta_i) = \sgn(\cos \theta_{j'}), \\
    \tilde{s}_2 = \sgn(\cos \theta_j) = \sgn(\cos \theta_{i'}).
  \end{aligned}
  \label{dualreflections}
\end{equation}
The use of these labels and the definition of $Q$ are shown geometrically in Figure \ref{fig3}.  One can think of $Q$ as the largest value of $q$ that one can obtain by 
reversing signs of $\tilde{s}_1$ and $\tilde{s}_2$.  For $\tilde{s}_1 = -\tilde{s}_2$, we have $Q=q$.  For $\tilde{s}_1 =  \tilde{s}_2$, however, $q$ becomes very small but $Q$ remains the same.  Summing over $\tilde{s}_1$ and $\tilde{s}_2$ is then required.


The absolute values in $\eqref{Qdefn}$ give four times the phase space as the previous expressions \eqref{qelt} and this overcounting needs to be adjusted for.  One factor of two can be explained by the fact that now 
$\vec q$ is no longer in fixed direction, since 
reversing the sign of both $\tilde{s}_1$ and $\tilde{s}_2$ 
changes $\vec q$ as $\vec{q} \to -\vec{q}$.  We therefore need to divide by $2$ to correct for this overcounting. 

The other factor of two arises because small $q$ processes are now described both by $X = Q/2k_F \sim \delta_T$ and $X \sim 1\gg \delta_T$. The latter case is what is referred to in \eqref{Qfromreflections}, and the former case corresponds to $Q = \min_{\tilde{s}_1, \tilde{s_2}} q$.  Both cases are a priori included in the change of variables \eqref{Qdefn}.  We
can discard the former case by only considering $X \gg \delta_T$, which both enables perturbation theory because now $X$ is always large and takes care of the overcounting problem. (We remind the reader that the goal of this procedure is to parameterize every collision so that perturbation theory done by expanding Chebyshev polynomials can be applied without breaking.)

%

One may verify explicitly that the entire integration range $\oint d\theta_i \oint d\theta_j$ is covered for $X \gg \delta_T$.
Indeed, for $s_1$ and $\tilde{s_1}$ equal to $1$, $\theta_i$ ranges from $0$ to $\pi/2$ as in \eqref{qelt}.  Then, 
reversing signs of $s_1$ and $\tilde{s_1}$ gives the other three quadrants.  The possible values of $\theta_j$ are similarly obtained by 
reversing signs of $s_2$ and $\tilde{s_2}$, where we note that some combinations such as $\theta_j = \pi/2$ and $\theta_i = 0$ are not attainable since they require $\omega \sim \varepsilon_F$ and that collisions with $\tilde{s_1} = \tilde{s_2}$ were previously hidden as small $q$ processes.  This is not an overcounting provided we only consider $X \gg \delta_T$.

A clean way to see that all possible $\theta_j$ consistent with $\theta_i$ are realized for $X \gg \delta_T$ is by noting that for $\tilde{s}_1 = - \tilde{s}_2$ we have a good understanding of all possible collisions via \eqref{qelt}.  Indeed, for these collisions either $x$ or $\tilde{x} = \tilde{q}/2k_F$ is much greater than $\delta_T$, and so the perturbation theory in \eqref{qangleexpand} can be used after a potential application of duality.  But 
sign reversal of $\tilde{s}_1$ and $\tilde{s}_2$ are bijective transformations that preserve the phase space measure, and hence all other processes are mapped one to one onto these. We can therefore include all allowed processes by summing over $\tilde{s}_1$ and $\tilde{s}_2$ while only considering $X \gg \delta_T$.

\begin{figure}[t]
  \includegraphics[width=0.99\columnwidth]{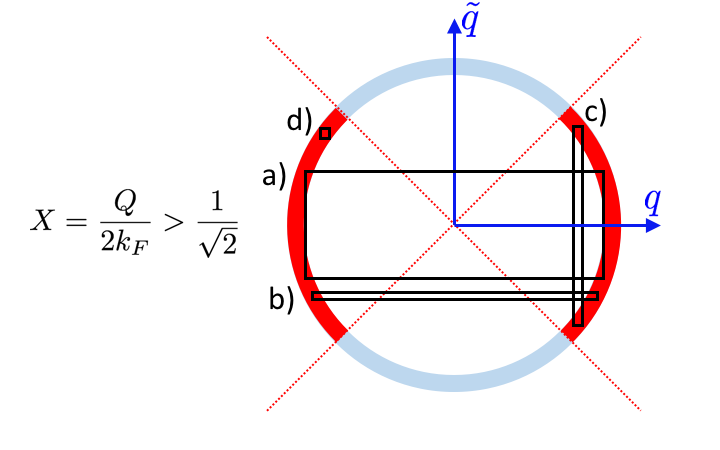}
\caption{Illustration of the condition $X > 1/\sqrt{2}$ and how it maps to $X < 1/\sqrt{2}$ under duality $q \leftrightarrow \tilde{q}$.  The rectangles labeled \textbf{a}, \textbf{b}, \textbf{c}, and \textbf{d} correspond to the corresponding collision types in Figure \ref{fig5}.
}
 \label{fig6}
\end{figure}

For convenience, we choose $X = Q/2k_F > 1/\sqrt{2}$.  Geometrically, this corresponds to fixing the domain of allowed $\theta_i$ and $\theta_j$ to within $\pi/4$ of $0$ or $\pi$, as depicted in Fig.\ref{fig6}.  We can do this consistently since for almost all processes if one of the momenta is in this region the others are automatically.  The exceptions are the rectangles where all momenta lie within $k_F T/T_F$ of the boundary, but these have a small phase space and can safely be ignored.  Since $X \mapsto \sqrt{1-X^2}$ under duality (taking $\theta \mapsto \theta - \pi/2$ in \eqref{Qdefn}), we are free to make such a restriction provided we multiply with an overall factor of $2$ that cancels out with the overcounting for $\pm q$.

It is possible to work out the Jacobian for the above transformation, 
\be\label{eq:J}
J=\frac{\p(\theta_i,\theta_j)}{\p(Q,w)}
,
\ee
but it is easier to drop the absolute values and demand consistency with \eqref{qelt}. Indeed, 
the differentiations in Eq.\eqref{eq:J} 
are unchanged under dropping absolute values and the above paragraphs show that the global factors due to potential overcounting are unchanged as well, provided we sum over $\tilde{s}_1, \tilde{s}_2$ and put the appropriate bounds on the $X$ integration.  In either case, and with the same approximations as in \eqref{jacobian}, we obtain
\begin{align}\nonumber
\oint d\theta_i \oint d\theta_j &  = \frac{T}{2m_*v^2}\sum_{s_1, s_2,  \tilde{s}_1, \tilde{s}_2} \int_{-\infty}^\infty dw
\\
& \times \int_{1/\sqrt{2}}^{1-\frac{a^2}{2}} \frac{dX}{X(1-X^2)}.
  \label{changeofvars}
\end{align}

We now need to expand $\sideset{ }{'}\sum_\alpha \cos m \theta_\alpha$ and $\sideset{ }{'}\sum_\alpha \sin m \theta_\alpha$.  The first expression is the same as before if $\tilde{s}_1 = - \tilde{s}_2$, but otherwise $\cos m \theta_j$ and $\cos m \theta_{i'}$ pick up a relative sign compared to $\cos m \theta_i$ and $\cos m \theta_{j'}$ and we obtain the combination $\pm(u_i + u_j - u_{i'} - u_{j'})$ which vanishes by energy conservation.  Therefore, for the $\cos m \theta_\alpha$ terms we reobtain \eqref{ustatechebyTq} and \eqref{1statechebyTq}
but with a lower $X$ limit of $1/\sqrt{2}$ and an upper $\theta$ limit of $\pi/4$ instead of $0$ and $\pi/2$ respectively.

It is straightforward to perform a similar computation with Chebyshev polynomials of the second kind for the $\sin m \theta_\alpha$ terms, but these are less convenient because of the $\sin \theta_\alpha$ prefactors and it is simpler and more illuminating to make use of duality.  In particular, taking the dual of the results from the previous paragraph, we obtain that the sine terms a) vanish for $s_1 = s_2$; b) yield a result proportional to $w$ instead of $\tilde{w}$; and c) are the same otherwise except with $\theta \mapsto \theta - \pi/2$.  The distinction between $w$ and $\tilde{w}$ is not important since their integrals against the Fermi functions are the same, and so we just replace $w$ with $\tilde{w}$ to match the cosine terms.

Putting everything together and rescaling energy variables with temperature, we obtain
\begin{equation}
  L_{\eta'\eta} = \frac{-m_* T^2}{32\pi^3\hbar v^2}\int d \mu_{iji'j'} \tilde{w}^2 \sum_{s_1,s_2,\tilde{s}_1, \tilde{s}_2}J_{\eta\eta'}
  \label{finalelt}
\end{equation}

where, for $\theta = \arccos X$,
\begin{widetext}
\begin{equation}
  \begin{aligned}
    J_{uu} & = \int_a^{\pi/4} \frac{d\theta}{\cos \theta \sin \theta }\abs{V}^2\lp\cos^2 m\theta\delta_{\tilde{s}_1, -\tilde{s}_2} + \sin^2 m\theta\delta_{s_1, -s_2}\rp,\\
     J_{1u} & = -\delta_T m\int_a^{\pi/4} \frac{d\theta}{\cos \theta \sin \theta}\abs{V}^2 (\cot \theta \delta_{\tilde{s}_1, -\tilde{s}_2}+  \tan \theta\delta_{s_1, -s_2}) \sin m\theta\cos m \theta, \\
     J_{11} & = \delta_T^2m^2 \int_a^{\pi/4} \frac{d\theta}{\cos\theta \sin\theta}\abs{V}^2 \left(\cot^2\theta  \sin^2 m\theta \delta_{\tilde{s}_1, -\tilde{s}_2} + \tan^2 \theta \cos^2 m \theta\delta_{s_1, -s_2} \right).
  \end{aligned}
  \label{Jelts}
\end{equation}
\end{widetext}
As in the previous evaluation, we can compute the leading order dependence on $m$ and $\log a^{-1}$ of the above.  The terms $\delta_{\tilde{s_1}, - \tilde{s_2}}$ and $\delta{s_1, -s_2}$ mean that in the limit of interest $\theta \to 0$ the only collisions that matter again look like the special head-on collision depicted in Fig.\ref{fig:specialprocess}b and have the corresponding matrix element $\abs{V_*}^2$.
\begin{equation}
  \begin{aligned}
    \sum_{s_1, s_2, \tilde{s}_1, \tilde{s}_2} J_{uu} & = 8\abs{V_*}^2\log\frac{1}{a} + \cdots, \\
    \sum_{s_1, s_2, \tilde{s}_1, \tilde{s}_2} J_{1u} & = -8\abs{V_*}^2\delta_T m^2W\log \frac{1}{ma} + \cdots, \\
    \sum_{s_1, s_2, \tilde{s}_1, \tilde{s}_2} J_{11} & = 8\abs{V_*}^2\delta_T^2 m^4W\log \frac{1}{ma} + \cdots.
  \end{aligned}
  \label{asymptoticqnew}
\end{equation}

We obtain the same result as \eqref{asymptoticq} except with an extra factor of two, as was anticipated, and argued, in the previous section. 
We therefore arrive at the final result:
\begin{equation}
  \gamma_m = \frac{\pi m_*^2 Wk_B}{15\hbar^5}\frac{T^4}{T_F^3}m^4 \log m.
\end{equation}
We have therefore repaired our earlier calculation by fixing all logical leaps and the the final prefactor.

\section{Conclusions}
\label{sec:Conclusions}
The long-lived collective excitations emerging out of momentum-conserving collisions in 2D Fermi gases is a surprising manifestation of  fermion exclusion. 
These excitations are of interest from a theory standpoint because they alter, in a fairly dramatic way, the traditional energy phase-space analysis of quasiparticle lifetimes. The excitation lifetimes that exceed the standard Fermi-liquid timescale by large factors of $(T_F/T)^2\gg 1$ suggest 
a range of theoretical and experimental implications. 

Besides exceptionally long lifetimes, the long-lived excitations have several other surprising properties. One is the distinct angular structure of an odd-parity modulation of the Fermi surface which protects these excitations from the dominant mechanism for angular relaxation in two dimensions: head-on collisions. Odd-parity excitations can only be relaxed through many small-angle collisions, and we find that this leads to relatively slow diffusion across the Fermi surface. Furthermore, this diffusion is not reduced to a simple Brownian random walk. Instead, it is dominated by correlated angular displacements of colliding particles, a process that leads to anomalous diffusion, or superdiffusion, described by a square of the Laplacian of the angular variable.


This physics defines a new transport regime that has a number of interesting experimental manifestations, of which we mention just a few.
One has to do with a beam of ``test particles'' injected into a two-dimensional Fermi gas. The dynamics of the beam  
will depend on collisional relaxation of its direction of motion.  In particular, head-on collisions will quickly give rise to a retroreflected hole beam that is observable by magnetically steering it into a nearby probe\cite{kendrick2018}.  At longer times, the forward electron beam and the backwards hole beam will slowly spread out through the anomalous diffusion we detailed above.

Another striking manifestation is that the existence of exceptionally long-lived modes 
alters the conventional ballistic-to-hydrodynamic crossover in 2D.  In particular, there will be an intermediate transport regime in which even-parity excitations have time to relax, but many odd-parity excitations do not.  This intermediate transport regime features non-local and scale-dependent conductivity and viscosity with nontrivial fractional power laws\cite{ledwith2017b}.  These fractional power laws are sensitive to the anomalous diffusion of the odd-parity excitations.

Looking ahead, the odd-parity modes can be expected to lead to interesting nonlinear effects in electron hydrodynamics. Indeed, the slow decay rates which make these modes long-lived will enhance the effects of nonlinearity. The reason for such enhancement is very general: because a long-lived mode, once activated, will be coupled to other modes during its lifetime, the net effect of nonlinearity will become stronger for longer lived modes. This opens up an exciting possibility to explore novel nonlinear effects and unconventional angular turbulence in driven electron systems.

We finally note that the picture discussed above has a considerable degree of universality. Namely, its validity is not limited to circular Fermi surface shape and parabolic band dispersion used in our analysis. Weak modulations of the Fermi surface, so long as they respect
inversion symmetry $\vec p\to -\vec p$, can be shown to preserve the unique role of head-on collisions 
and anomalously slow relaxation rates for the odd-parity harmonics. Parabolic band dispersion, likewise, is inessential at $T\ll T_F$, since near the Fermi level, where all the action is happening, a nonparabolic band can always be approximated by a parabola with curvature set by the effective mass. Disorder and Umklapp scattering, on the other hand, can present a limitation, however these effects are weak in modern 2D materials such as graphene and GaAs-based electron systems, where the new physics due to long-lived odd-parity modes can be realized and explored.
%
%
%
%
%

 \renewcommand{\theequation}{A-\arabic{equation}}
  \setcounter{equation}{0}  

  \section*{Appendix: Some useful Fermi integrals}  
\label{appendix:integrals}
Here we consider integrals of the type
\begin{equation}
    \langle P(u_\alpha)\rangle\equiv\int d\mu_{ji'j'} P(u_\alpha),
\end{equation}
 over the measure
\be
\begin{split}
d\mu_{ji'j'}&= \frac{du_j du_{i'}du_{j'}}{f_i(1-f_i)} \delta(u_i+u_j-u_{i'}-u_{j'}) \\
&\times f_if_j(1-f_{i'})(1-f_{j'}),
\end{split}
\ee
where $P(u_\alpha)$ is a general polynomial in $u_j,u_{i'}$ and $u_{j'}$ and $f(u)=1/(e^u+1)$. The above integrals are used in Sec.\ref{sec:perturbation_theory} and Sec.\ref{sec:matrix_element_evaluation}.

To obtain $\langle P(u_\alpha)\rangle$, we consider the generating functional
\begin{equation}
    J[\beta_a]\equiv \left\langle \exp(i\beta_j u_j-i\beta_{i'}u_{i'}-i\beta_{j'}u_{j'})\right\rangle.
\end{equation} 
where $\beta_a$, $a=j,i',j'$ are auxiliary parameters.

 To 
 evaluate $J[\beta_a]$ we proceed in three steps. First, we make a simple variable change
\be
u_{i'}\to -u_{i'}
,\quad
u_{j'}\to -u_{j'}
\ee
and use the property of Fermi functions $f(-u)=1-f(u)$ to transform the measure to a symmetric form (the sum goes over $a=j,j',i'$)
\be
J= \int e^{i\sum_a \beta_a u_a}\frac{du_j du_{i'}du_{j'}}{1-f_i} \delta(u_i+u_j+u_{i'}+u_{j'})  f_jf_{i'}f_{j'}.
\ee
Next, using the identity $\delta(x)=\int_{-\infty}^\infty \frac{d\alpha}{2\pi} e^{-i\alpha x}$ with $x=u_i+u_j+u_{i'}+u_{j'}$, we rewrite this expression as
\be
J=\int_{-\infty}^\infty \frac{d\alpha}{2\pi}\frac{e^{-i\alpha u_i}}{1-f_i} \prod_{a=j,i',j'}
 \int_{-\infty}^\infty du e^{i(\beta_a-\alpha) u} f(u)
.
\ee
This, combined with the known Fourier transform
\[\int_{-\infty}^\infty du e^{-i(\alpha +i\delta)u} f(u)=\frac{\pi i}{\sinh[(\alpha+i\delta)\pi ]}, \] with an infinitesimal imaginary part added to assure convergence, yields
\be
J=\int_{-\infty}^\infty \frac{d\alpha}{2\pi}\frac{e^{-i\alpha u_i}}{1-f_i} \prod_{a=j,i',j'}
\frac{\pi i}{\sinh[\pi(\alpha-\beta_a+i\delta)]}.
\ee
This procedure reduces the original three-dimensional integral to a one dimensional integral.

The integral over $\alpha$ can be computed by noting 
that under a shift $\alpha\to\alpha-i$ the integrand picks up a factor $-e^{-u_i}$, so the integral can be computed using a rectangular contour $C$ 
encircling the strip $-1<\Im\alpha<0$, with the sides on the lines
$\Im\alpha=0$ and $\Im\alpha=-1$. The contributions from $\Im\alpha=0$ and $\Im\alpha=1$ together cancel the $1-f_i$ factor. Within the contour, the integrand has three poles at $\alpha=\beta_a-i\delta$, $a=j,i',j'$; evaluating residues at the poles, we have
\begin{equation}
\begin{split}
J&=\frac{-\pi^2 e^{-i\beta_j u_i}}{\sinh[\pi(\beta_j-\beta_{i'})]\sinh[\pi(\beta_j-\beta_{j'})]}\\
&+\frac{-\pi^2 e^{-i\beta_{i'} u_i}}{\sinh[\pi(\beta_{i'}-\beta_{j})]\sinh[\pi(\beta_{i'}-\beta_{j'})]}\\
&+\frac{-\pi^2 e^{-i\beta_{j'} u_i}}{\sinh[\pi(\beta_{j'}-\beta_{i'})]\sinh[\pi(\beta_{j'}-\beta_{j})]}.
\end{split}
\end{equation}
Differentiating $J$ with respect to $\beta_a$'s, we can obtain various integrals used in Secs.\ref{sec:perturbation_theory} and \ref{sec:matrix_element_evaluation}:
\begin{eqnarray}
\int d\mu_{ji'j'}=&\frac{\pi^2+u_i^2}{2},\\
\int d\mu_{ji'j'}u_j=&-u_i \frac{\pi^2+u_i^2}{6},\\
\int d\mu_{ji'j'}(u_i-u_j)^2=&\lp\frac{\pi^2+u_i^2}{2}\rp^2,
\end{eqnarray}
and
\be
\begin{split}
    \int d\mu_{iji'j'} (u_i-u_j)^2&=\int du_i f_i(1-f_i)\lp\frac{\pi^2+u_i^2}{2}\rp^2\\
           & =\frac{8\pi^4}{15}.
\end{split}
\ee
This provides a derivation of the results used in Eqs.\eqref{L1|1>} and \eqref{eq:8pi4/15}.


\end{document}